\documentclass[aps,prd,twocolumn,amsmath,amssymb,showpacs,floatfix,nofootinbib]{revtex4-1}

\usepackage{graphicx}
\usepackage{dcolumn}
\usepackage{xcolor}
\usepackage{bm}
\usepackage{natbib}
\usepackage{calligra}
\usepackage[T1]{fontenc}
\usepackage{egothic}
\usepackage[T1]{fontenc}
\newfont{\rsfsten}{rsfs10 scaled 1200}
\newfont{\rsfsseven}{rsfs10 scaled 1200}
\newfont{\rsfsfive}{rsfs10 scaled 1200}
\usepackage{epsfig}
\usepackage{units}
\usepackage[utf8]{inputenc}
\usepackage{multirow}

\newcommand{\aap}{{Astron.~Astrophys.}}

\newcommand{\aj}{{Astron.~J.}}
\newcommand{\mnras}{{Mon.~Not.~R.~Astron.~Soc.}}

\advance \voffset by -0.5cm\relax

\def\lsim{\mathrel{\raise.3ex\hbox{$<$\kern-.75em\lower1ex\hbox{$\sim$}}}}
\def\gsim{\mathrel{\raise.3ex\hbox{$>$\kern-.75em\lower1ex\hbox{$\sim$}}}}

\begin{document}

\title{Limits on Runaway Growth of Intermediate Mass Black Holes from Advanced LIGO}

\author{Ely D. Kovetz}
\author{Ilias Cholis}
\author{Marc Kamionkowski}
\author{Joseph Silk}
\affiliation{\vspace{0.05in}Department of Physics and Astronomy, Johns Hopkins University, Baltimore, MD 21218 USA}

\begin{abstract}
There is growing evidence that intermediate-mass black holes (IMBHs), defined here as having a mass in the range $M=500-10^5\,M_\odot$, are present in the dense centers of certain globular clusters (GCs). Gravitational waves (GWs) from their mergers with other IMBHs or with stellar BHs in the cluster are mostly emitted in frequencies $\lesssim10\,{\rm Hz}$, which unfortunately is out of reach for current ground-based observatories such as advanced LIGO (aLIGO). 
Nevertheless, we show that aLIGO measurements can be used to efficiently probe one of the possible formation mechanisms of IMBHs in GCs, namely a runaway merger process of stellar seed BHs. In this case, aLIGO will  be sensitive to the lower-mass rungs of the merger ladder, ranging from the seed BH mass to masses $\gtrsim50-300\,M_\odot$, where the background from standard mergers is expected to be very low.  Assuming this generic IMBH formation scenario, we calculate the mass functions that correspond to the limiting cases of possible merger trees. Based on estimates for the number density of GCs and taking into account the instrumental sensitivity, we show that current observations do not effectively limit the occupation fraction $f_{\rm occ}$ of IMBHs formed by runaway mergers of stellar BHs in GCs. However, we find that a six-year run of aLIGO at design sensitivity will be able to probe down to $f_{\rm occ}\lesssim3\%$ at a 99.9\% confidence level, either finding evidence for this formation mechanism, or necessitating others if the fraction of GCs that harbor IMBHs is higher.
\end{abstract}

\keywords{binaries: close --- stars: evolution,....}

\maketitle

\section{Introduction}

Black holes have so far been detected in two separate mass ranges. One is the supermassive
black hole (SMBH) range, roughly $10^5-10^9\,M_\odot$. SMBHs have been indirectly observed 
throughout the cosmos, from the center of the Milky Way (by studying the kinematics of stars near the 
Galactic center \cite{Mortlock:2011va}) and up to redshifts of more than 7 (detecting emission at various wavelengths 
originating from the centers of their host galaxies \cite{Ghez:1998ph, Ghez:2008ms}). 
The other is the stellar-mass range, from a few to a few tens of solar masses, where 
BHs were first detected through the observation of X-rays emitted as a result of their accretion 
from a binary partner \cite{Motch:1996gw, in'tZand:2000zz, Grimm:2001vd, Lutovinov:2004wi, Corbet:2007vn, 
Russell:2013jva,Corral-Santana:2015fud, Bogomazov:2016cei}, and more recently have been detected by measuring gravitational 
waves originating from BH binary coalescences \cite{Abbott:2016blz,Abbott:2016nmj, TheLIGOScientific:2016pea}. 
An interesting regime left to explore is the intermediate range, which we take here to be $\sim500-10^5\,M_\odot$.  
The lower  mass limit is chosen in order to avoid confusion with ``second generation'' BHs \cite{Fishbach:2017dwv,Gerosa:2017kvu}---the 
product of single mergers of stellar BHs---which can be as massive as $\simeq 100 M_{\odot}$.
  
There are various indications, albeit mostly circumstantial, for the existence of 
such intermediate-mass black holes (IMBHs) \cite{Miller:2003sc}.
Some empirical evidence comes from the observation of Ultra-Luminous X-ray sources, 
with luminosities as high as $10^{40}\,{\rm ergs/sec}$,  larger than the Eddington
luminosity of stellar-mass BHs, hinting towards more massive accreting objects 
\cite{Berghea:2008rc,2009Natur.460...73F,2011ApJ...734..111D,Godet:2014bga,Pasham:2015tca}. 
Evidence of ULX variability, especially for sources in M82 \cite{Bachetti:2014qsa}, 
suggests some may be rapidly accreting highly super-Eddington neutron stars, based on periodicity, 
while others are quasi-periodic oscillations that are most likely accreting IMBHs of 
a few hundred solar masses \cite{Bachetti:2015pwa}. While one could possibly explain this 
by considering a cluster of small sources instead of an IMBH, the $O(10)$ variability of the radiation over periods 
of months makes this unlikely \cite{Mezcua}. In addition, the sources are
often detected away from the host galaxy centers \cite{Miller:2003sc}, and are therefore unlikely to be due to 
SMBHs. Moreover, the inferred masses of these sources are consistent 
with an extrapolation of the $M_{\rm BH}-\sigma$ relation, which holds across decades of masses 
in the SMBH range \cite{2013ApJ...764..151G}. Recently, indirect evidence was reported for the presence of IMBHs in 
NGC 104---a globular cluster (GC) we shall investigate more closely below---based on dynamical analysis of its pulsars \cite{Kiziltan} (see however \cite{Freire:2017mgu}), as well as for NGC 6624 \cite{Perera:2017jrk}.
    
Theoretical arguments also support the existence of IMBHs, 
 with an occupation fraction as high as unity in dwarf galaxies (as early feedback they induce can provide a solution
 to a number of pressing dwarf-galaxy anomalies \cite{Silk:2017yai}), as well as in GCs (where they 
 can potentially account for the ``missing link'' in generating SMBHs at high redshift \cite{Ebisuzaki:2001qm}).

Advanced LIGO (aLIGO) has already detected a series of stellar mass black hole mergers in its 
first two observing runs, with pre-merger masses as massive as $\gsim30\,M_\odot$, and as its  
sensitivity sharply increases for higher masses, one might hope that IMBHs would pose an easy target for detection.
However, IMBHs lie in a problematic regime for aLIGO. To detect a merger involving an IMBH, 
we could either hope to see it merging with another IMBH, or with abundant 
compact objects in the host GC (such as neutron stars or stellar-mass BHs). For the 
former case, the signal from its inspiral and merger phases lies outside the frequency 
range of aLIGO, if $M_{\rm IMBH}\gtrsim500\,M_\odot$. 
The latter case is termed an intermediate mass-ratio inspiral (IMRI).
While it may be accessible at slightly higher frequencies, the exact waveforms are hard to model \cite{Brown:2006pj,Smith:2013mfa}, 
as the post-Newtonian approximation---an expansion in $v/c$---fails when the velocity 
approaches the speed of light, a regime win which IMRIs spend many cycles \cite{Mandel:2008bc}. Thus accurate
templates are hard to calculate\footnote{Note that the case for extreme mass-ratio inspirals (EMRIs) is 
somewhat simpler, as linearization based on a perturbative expansion in $m/M$ is valid for mass 
ratios $>O(10^6)$ \cite{AmaroSeoane:2007aw,Gair:2010yu}.}. It is worth noting that a quantum of solace, so to speak, lies
in the fact that the ringdown process following massive mergers includes non-negligible 
GW emission at higher frequencies (the ``quasi-normal modes'') \cite{Flanagan:1997sx,Berti:2005ys}, some within the LIGO range \cite{Fregeau:2006yz,Shinkai:2016xya}.
This prevents its detectable volume from shrinking quickly to zero as the total mass approaches the
IMBH range \cite{Abbott:2017iws,Fishbach:2017zga,CalderonBustillo:2017skv}. 
 
Rather than try to observe IMBHs directly, for which we likely have to wait until the 2030s 
(or later), when next-generation experiments 
can more easily access IMBH-IMBH mergers or EIMRIs \cite{Gair:2010dx},
in this paper we take an alternative approach. We focus on the prospects of using existing and 
upcoming results from aLIGO to probe the {\it formation} of these IMBHs. 
  
The formation process of IMBHs is highly uncertain. Several scenarios have hitherto been suggested,
for example describing how massive BH seeds can be formed from direct collapse of Population III stars \cite{Miller:2003sc} or following consecutive 
mergers of stars in young dense stellar clusters \cite{Ebisuzaki:2001qm,AtakanGurkan:2003hm,PortegiesZwart:2004ggg,Sakurai:2017opi}. 
An obvious route to IMBH formation is repeated mergers of stellar remnants such as BHs in (old) dense clusters, 
which can be efficient if a non-linear runaway process can take place \cite{Miller:2001ez}. 
Over the past decade, numerical simulations of such mergers in GCs, showing that
most tight binaries are ejected from the cluster before or after merging, have cast heavy doubts on this scenario \cite{PortegiesZwart:1999nm,Rodriguez:2015oxa}. 
However, recent works taking into account higher-order post-Newtonian dynamics have shown both analytically \cite{Samsing:2017xmd}
and numerically \cite{Gondan:2017wzd} that significantly(!) more (by up to two orders of magnitude) binaries are retained and 
as much as half of the total binaries can merge within the cluster, reviving the scenario and motivating attempts to test it observationally.

Fortunately, if indeed an IMBH is formed by a runaway merger starting from small seed BHs, aLIGO may be 
sensitive enough to detect the
early stages of this process. We show that provided the fraction of GCs hosting such IMBHs 
is non-negligible, aLIGO will detect at a minimum several mergers per year involving BHs with masses 
larger than $100\,M_{\odot}$ resulting from their formation.
Alternatively, as our analysis demonstrates, a null detection of mergers involving $
\sim50-300\,M_\odot$ black holes by aLIGO running at design sensitivity can place 
useful limits on $f_{\rm occ}$, the occupation fraction of IMBHs formed by 
a runaway merger of stellar black holes in GCs. Our conclusion is that with six years of aLIGO running at design
sensitivity, using the most conservative assumptions about the signal and background, 
aLIGO will be able to robustly probe down to $f_{\rm occ}\lesssim3\%$ at 99.9\% confidence.

This paper is constructed as follows: 
In Section~\ref{sec:setup} we describe the setup for our analysis and introduce our target observable, 
the total observed mass distribution of merging BHs by aLIGO. This includes contributions from standard 
mergers of stellar BHs, which form a {\it background}, and those stemming from runaway mergers 
leading up to IMBH formation in GCs, which is the {\it signal} we are after. In Section~\ref{sec:Signal} we 
model the signal, considering opposite limiting cases for the possible merger trees leading up to the IMBH and using a simple
ansatz for the rate of such events and its redshift dependence. In Section~\ref{sec:Expectation} we derive an expected 
value for $f_{\rm occ}$, based on analytic estimates of the rates of binary capture and merger in the centers of GCs, which
we calculate for a representative sample of Milky-Way GCs. In Section~\ref{sec:Background}, we account for 
the BH merger events not associated with the runaway process that act as a background to its observation, following 
the prescription in Ref.~\cite{Kovetz:2016kpi}.  
In Section~\ref{sec:LIGOSensitivity}, we present our calculations for the detectable spacetime volume that aLIGO 
has access to for mergers of BHs of given masses. Our results are  
presented and explained in Section \ref{sec:results}. 
We conclude in Section \ref{sec:Conclusions}.


\section{Setup}
\label{sec:setup}
  
In a nutshell, we wish to place a limit on the abundance of IMBHs in GCs that were formed by a runaway merger of smaller
 black holes by studying the mass distribution of merging BHs detected by aLIGO. We leave a discussion of the spin distribution,
 which may provide an additional  means of probing hierarchical mergers \cite{Fishbach:2017dwv}, to future work.
 The number of observed mergers with a heavier component of mass $M$ is given by
\begin{equation}
N_{\rm obs}(M)=\int\limits^{M+\Delta M}_{M-\Delta M}\frac{dn}{dM}\bar{R}\langle VT\rangle dM,
\label{eq:Nobs}
\end{equation}
where $dn/dM$ is the mass function of merging black holes, $\bar{R}$ is the overall event rate, which in general can be redshift dependent, 
and $\langle VT\rangle$ is the space-time volume detectable by aLIGO. 
To calculate this observable, we distinguish between the {\it signal} coming from the formation of IMBHs and the 
{\it background}, which is due to mergers occurring otherwise throughout the Universe.

The signal contribution depends (linearly) on $f_{\rm occ}$, the occupation fraction of runaway-formed IMBHs in GCs. 
After we present our prescription for calculating the signal in the next Section, we will discuss what is the expected value for this quantity,
to be later compared with our final results. Next we will review the formalism presented in Ref.~\cite{Kovetz:2016kpi} for calculating the background contribution.
We will then describe our use of the LIGO Scientific Collaboration Algorithm Library Suite (LALSuite) \cite{LALwebsite} to calculate the sensitive volume of aLIGO to mergers of given masses with noise levels corresponding to its completed O1 and O2 runs, its upcoming O3 run in late 2018 and the final run at design sensitivity planned for 2020-2025. 

At this point, armed with a prediction for $N^{\rm Signal}_{\rm obs}(M)+N^{\rm BG}_{\rm obs}(M)$ where $N^{\rm Signal}_{\rm obs}(M)\propto f_{\rm occ}$, 
we will be in a position to derive an upper bound on $f_{\rm occ}$ based on the null hypothesis that there are no detectable events originating from 
runaway mergers. Since the expected number of background events at high masses is low, we will have to go beyond the Gaussian approximation for estimating the significance of a peak in an individual mass bin and use Poisson statistics to calculate a $99.9\%$-confidence upper bound on $f_{\rm occ}$. To achieve robust conclusions, we will repeat our calculations for wide a range of models (and their free parameters) for the background distribution.


\section{Signal}
\label{sec:Signal}

Referring to Eq.~(\ref{eq:Nobs}), in order to calculate the signal we are after, we need to estimate $dn/dM$, the mass distribution of merging black holes in the runaway process, and $\bar{R}$, the overall rate of IMBH formation in GCs.

\subsection{Runaway Merger Trees}

\begin{figure*}
\begin{centering}
\includegraphics[width=\columnwidth]{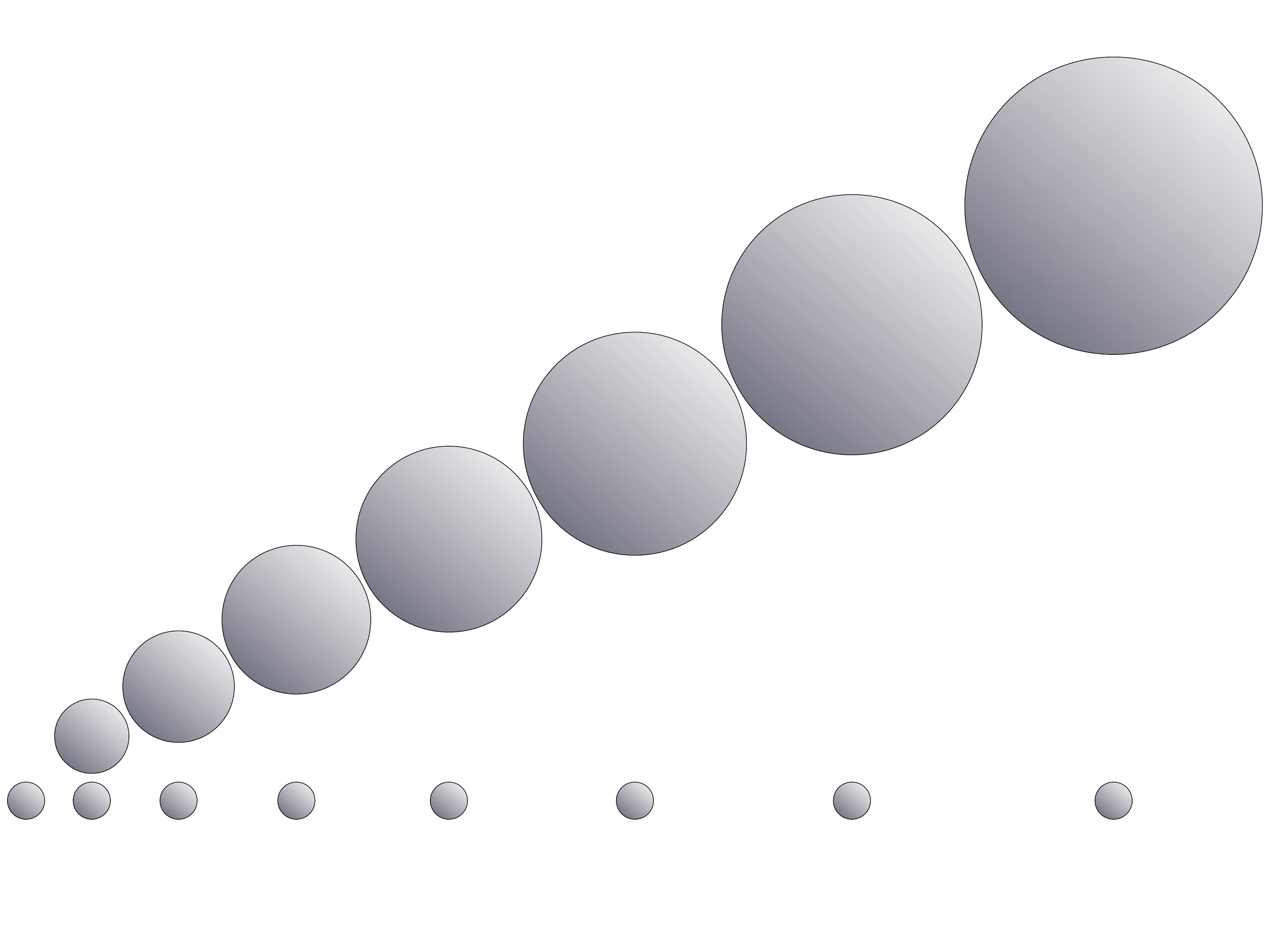}
\includegraphics[width=\columnwidth]{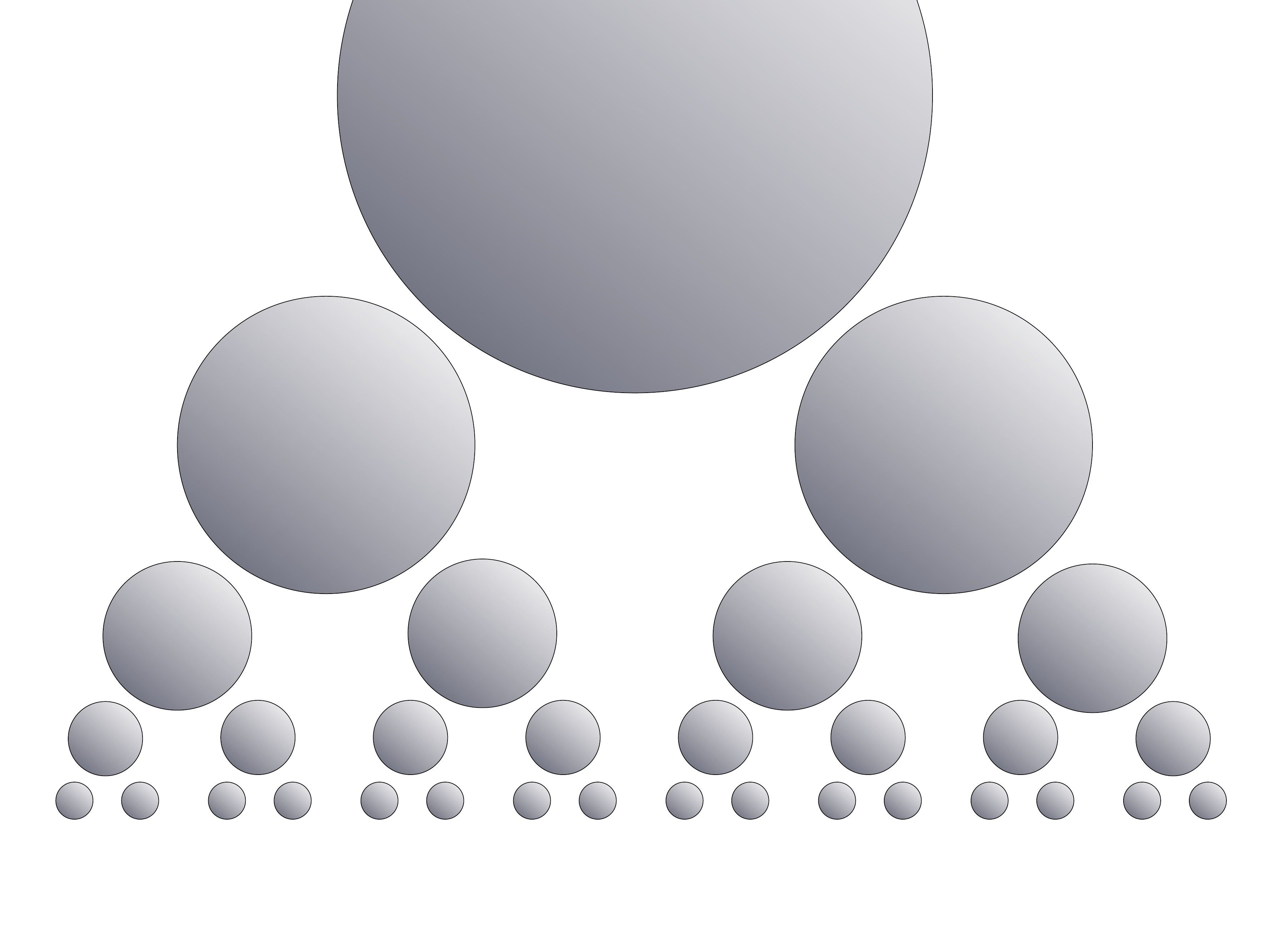}
\end{centering}
\caption{{\it Left:} The top-heavy ``Right Triangle'' scenario, where each merger involves the most massive BH and a $10\,M_\odot$ BH. 
{\it Right:} The bottom-heavy ``Isosceles'' scenario, where all mergers are of equal-mass binaries, producing more detectable events.} 
\label{fig:IMBHGrowth}
\end{figure*}

Without loss of generality, as this is trivial to extend, our setup assumes that globular clusters across the Universe are populated 
with seed BHs with mass $M_{\rm seed}=10\,M_\odot$. These can then repeatedly merge and follow different paths towards generating an IMBH.
The mass distribution of mergers in the runaway process depends on the precise path leading up to the formation of an IMBH. 
Naturally, the actual merger tree for each and every IMBH cannot be determined. Therefore, we shall consider the limiting cases \cite{Matsubayashi}, 
illustrated in Fig.~\ref{fig:IMBHGrowth}, and estimate the resulting mass dependence of the merger rates.
  
The bottom-heaviest route to growing an IMBH that can be achieved is when mergers of equal-mass black holes are 
dominant and the hierarchy of mass mergers quadratically approaches the final mass, as illustrated in the right panel 
of Fig.~\ref{fig:IMBHGrowth}. Assuming the bottom of the mass range is made up of equal mass seed BHs, 
it is easy to see that the number of mergers scales with the merging masses as $n(M)\propto1/M$ \cite{Shinkai:2016xya}. 
Therefore, the mass function $dn/dM$ of the merging BHs will be given by
\begin{equation}
\frac{dn}{dM}(M)_{\rm bottom}= C/M^{2},
\label{eq:MFBottom}
\end{equation}
where the normalization ensures that integrating over the mass of the merging BHs from $M_{\rm seed}$ to  half 
the IMBH mass we get the correct total number of mergers,
\begin{eqnarray}
&&N_{\rm mergers}~~=\int\limits^{M_{\rm IMBH}/2}_{M_{\rm seed}}\frac{C}{M^2}dM ~~\equiv~~ M_{\rm IMBH}/M_{\rm seed}-1 \nonumber \\
&& \longrightarrow C=\left(\frac{M_{\rm IMBH}}{M_{\rm seed}}-1\right)/ \left(\frac{1}{M_{\rm seed}}-\frac{2}{M_{\rm IMBH}} \right)\sim M_{\rm IMBH}. \nonumber \\
\end{eqnarray}

Meanwhile, the opposite top-heavy route to creating an IMBH is when mergers with the highest mass-ratio are dominant, as shown in the left 
panel of Fig.~\ref{fig:IMBHGrowth}. Here, the number of mergers is $n(M)\propto M$. Since $M_{\rm seed}$ is fixed by construction, the total number of mergers is the 
same as above, and similarly we can derive the mass function,
\begin{eqnarray}
\frac{dn}{dM}(M)_{\rm top}= \left(\frac{M_{\rm IMBH}}{M_{\rm seed}}-1\right)/(M_{\rm IMBH}-2M_{\rm seed})\sim\frac{1}{M_{\rm seed}}. \nonumber \\
\label{eq:MFTop}
\end{eqnarray}

\subsection{The overall rate of the runaway process}
\label{sec:TotalMR}
  
Next, we need to determine how often the runaway process occurs in GCs anywhere in the local Universe.
We assume that the local GC density is $n_{\rm GC} \simeq3\,{\rm Mpc}^{-3}$ \cite{PortegiesZwart:1999nm} and that a fraction $f_{\rm occ}$ of these 
GC centers are occupied by an IMBH with mass $M_{\rm IMBH}$ that has been generated from
 $M_{\rm IMBH}/M_{\rm seed}-1$ mergers starting with $10\,M_\odot$ seed BHs over the lifetime $t_{\rm age}$ of the cluster.
 The latter has to be less than a Hubble time, and we set it at $t_{\rm age}=10\,{\rm Gyr}$, to allow some time for the formation 
 of the cluster, its BHs, their relaxation and segregation towards the center, etc.~(we discuss these timescales further in Section~\ref{sec:Expectation}).
The overall IMBH formation rate is then simply given by
\begin{equation}
\bar{R} = f_{\rm occ}n_{\rm GC}/t_{\rm age} = 0.3 f_{\rm occ}\,\, {\rm Gpc^{-3} yr^{-1}}.
\label{eq:signalrate}
\end{equation}
Note that we assume that the overall rate of the runaway process in GCs is independent of the 
merger trees discussed above. This may not be true in all cases, but
we leave that discussion for future work.  
We also note that we expect an occupation fraction $f_{\rm occ}$ much less than unity, a point which we discuss in detail in 
Section~\ref{sec:Expectation}.

\subsection{Redshift Dependence}

The rate in Eq.~(\ref{eq:signalrate}) is redshift independent. While the simplest case to consider is when both 
the signal and background are assumed not to evolve with time, it is instructive to compare with what happens 
when we adopt a (different) redshift dependence for both. To account for time evolution of the rate of IMBH formation, 
we will simply follow Ref.~\cite{Fragione:2017blf} and replace the fixed GC number density above with $n_{\rm GC}(z)\propto [H(z)/H_0]^3$
(to avoid including additional uncertain assumptions, we do not correct the locally measured $n_{\rm GC}(0)$ to account 
for how many primordial GCs  survive until a given redshift $z$).


\section{Expectation}
\label{sec:Expectation}

The goal of this Section is to derive a reasonable expectation for the range of values the quantity we focus on in this work---the occupation fraction $f_{\rm occ}$ of runaway-formed IMBHs in GC centers---can take. We do this based on analytic estimates of the BH merger rates due to various binary capture mechanisms, calculated for a representative sample of observed Milky-Way (MW) GCs. 

\subsection{Characteristic sample of GCs}

\begin{figure}
\begin{centering}
\includegraphics[width=\columnwidth]{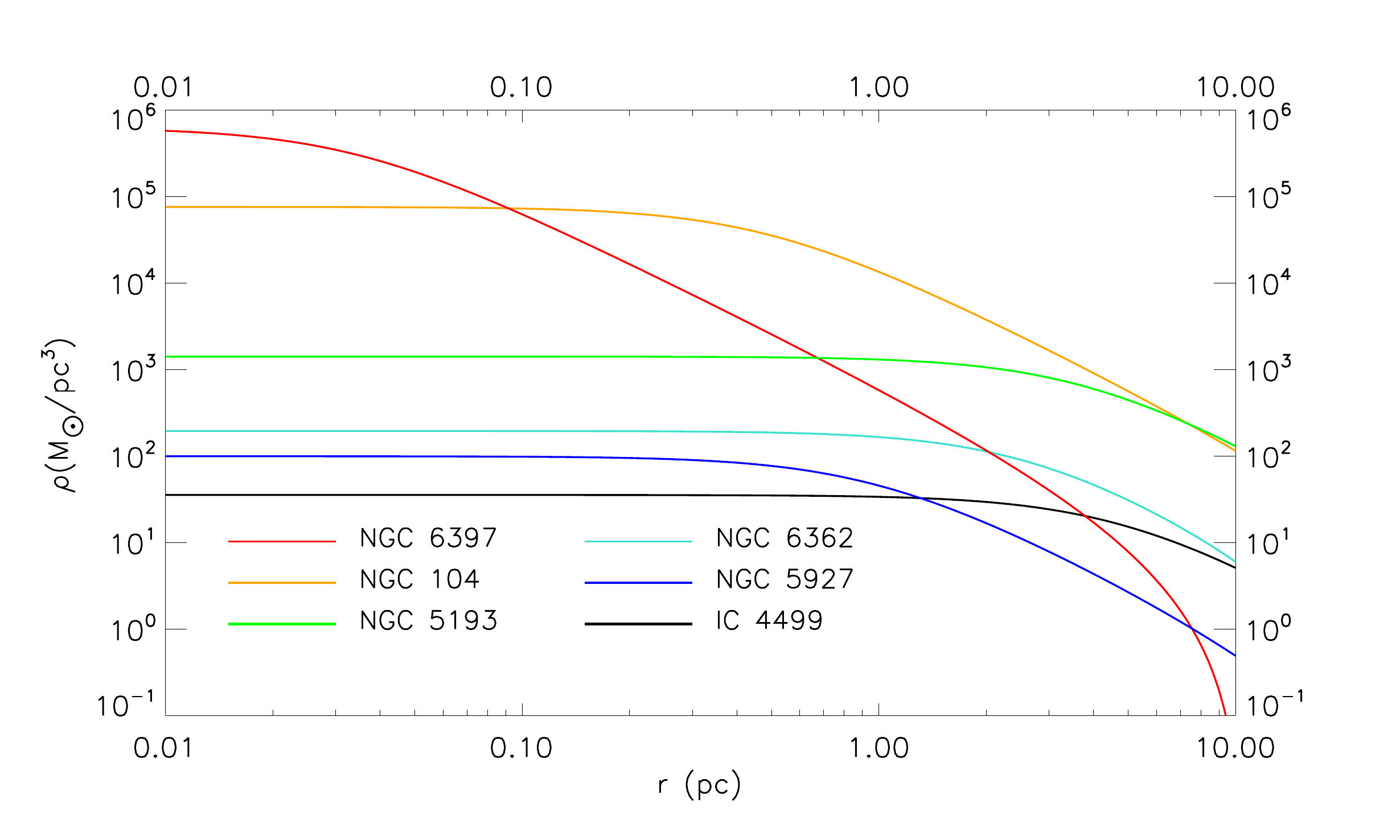}
\end{centering}
\vspace{-0.2in}
\caption{The density profiles of six Milky-Way globular clusters, 
spanning a wide range of properties.}
\label{fig:GCProfiles}
\end{figure}

There are hundreds of GCs detected to date. Using data from \cite{NASAwebsite} as observational input, 
we present in Fig.~\ref{fig:GCProfiles} the density profiles of six GCs in the MW, assuming that their total stellar 
mass follows the King profile \cite{King1962},
\begin{equation}
\rho(r) \propto \left[\left(1 + (\frac{r}{r_{c}})^2\right)^{-1/2} -  \left(1 + (\frac{r_{t}}{r_{c}})^2\right)^{-1/2} \right]^{2},
\end{equation}
where $r_{c}$ is the core radius and $r_{t}$ the tidal radius of the cluster. 
This sample spans a wide range of GC properties, from very dense (NGC 6397) or massive (NGC 104 and NGC 5139) to 
very sparse and light (NGC 5927). For the remainder of this section, we will focus on three of these clusters as representative cases: 
NGC 104, NGC 6362 and NGC 5927, and discuss the dynamical evolution of BH stellar remnants in such environments to determine 
which GCs can host IMBHs \cite{Baumgardt:2004tp}. We note
that a recent analysis of pulsar accelerations in NGC 104 yields evidence for the existence of a $\sim{2300\,M_\odot}^{+1500}_{-850}$ at its center \cite{Kiziltan} (though this finding has been challenged by Ref.~\cite{Freire:2017mgu}).
  
In Table~\ref{tab:NGC}, we show the central mass densities (inferred from the observed central luminosity density), the measured velocity 
dispersion, and the estimated maximum mass in BHs (taken to be $10\,M_\odot$ each, consistent with our choice throughout this work) in these GCs. 
We calculate the latter using the Kroupa mass function \cite{Kroupa:2000iv} and assuming that all stars with initial mass $> 25 M_{\odot}$ become BHs (realistically, less than $20 \%$ of these BHs---and maybe only a few $\%$ in the smaller clusters---will remain dynamically bound in the GCs after the natal kick).
\begin{table}[!ht]
\begin{center}
\begin{tabular}{c|ccc} \hline
GC Name &$\rho(r=0)~ [M_{\odot}/\textrm{pc}^3] $&$v_{\textrm{DM}}~ {[\rm km/sec]}$& $N_{\textrm{BH}}^{\textrm{max}}$\\   \hline
NGC 104 &$7.6 \times 10^{4}$ & 11 &$4.4 \times 10^{4}$ \\ 
NGC 6362 & $2.0 \times 10^{2}$ & 2.8 &$1.5 \times 10^{3}$ \\ 
NGC 5927 & $1.0 \times 10^{2}$ & 2.5 &$1.5 \times 10^{2}$ \\ 
\hline
\end{tabular}
\end{center}
\caption{The characteristics of three GCs. $\rho(r=0)$ refers to the mass density at the center of the GC. $N_{\textrm{BH}}^{\textrm{max}}$ 
is the maximum possible number of $10\,M_{\odot}$ seed BH that can exist in these GCs (see main text). $v_{\rm DM}$ is the velocity dispersion.}
\label{tab:NGC}
\end{table}

Before we turn to calculating the rates for binary capture, we first estimate the timescales for the relevant processes in the earlier stages of the GC evolution.
The first timescale to consider  is that for BH formation. The most massive objects in the GCs, O-type stars that will give birth to BHs after core-collapse, 
have typical lifetimes $\lsim 10\, {\rm Myrs}$. 
The second is the relaxation timescale of an object of mass $m$ in a GC, which is given by,
\begin{equation}
\tau_{\textrm{relax}} = \frac{2 \times 10^{12}\, \textrm{yr}}{\ln \left(\frac{R_{\textrm{max}}}{R_{\textrm{min}}}\right)} 
\left( \frac{v}{10\, {\rm km/s}} \right)^{3} \left( \frac{M_{\odot}}{m} \right)^{2} \left( \frac{1\, {\rm pc}^{-3}}{n} \right), 
\label{eq:tauRelax}
\end{equation}
where $v$ is the velocity relative to the stars, $n$ the stellar density, $R_{\textrm{max}}$ the size of the
system (i.e.\ the GC) and $R_{\textrm{min}} \sim {\rm au}$ the distance where a strong encounter with another star will take place. 
Our choice of BH seed mass sets $m = 10\, M_{\odot}$. Including all relevant 
numbers, we get that a seed BH in NGC 104 will have a relaxation timescale of $10^{5}\, {\rm yrs}$, while for NGC 6362 
it is $\simeq 3 \times 10^{5}\,{\rm yrs}$ and for NGC 5927 that grows only to $\sim 10^{6}\, {\rm yrs}$. These  are 
shorter than the lifetimes of the progenitor stars.

Finally, the last  timescale relevant here is that of dynamical friction, which for an object of mass $m$ in a 
cluster with density $\rho$ and velocity dispersion $v_{\textrm{DM}}$ scales as,
\begin{equation}
\tau_{DF} = \Lambda \frac{v_{\textrm{DM}}^{3}}{4 \pi G^{2} \rho m},
\label{eq:tauDF}
\end{equation}
where $\Lambda \sim O(1)$. Substituting again the characteristic values from Table~\ref{tab:NGC}, we get that for the $10\,M_{\odot}$
BHs that reside within the cores of the clusters, the dynamical friction timescale is $20-100\,{\rm Myrs}$.  
The combination of the dynamical friction and relaxation timescales thus leads us to the conclusion that within $\simeq 100\,
{\rm Myrs}$ the seed BHs have segregated in the center of the clusters (regardless of the exact cluster densities and 
masses). 

\subsection{Rates of relevant capture mechanisms}
\label{subsec:GCMR}

There are several mechanisms that can contribute to mergers involving BHs.
These include the direct gravitational-wave capture of two BHs; three-body effects relevant especially for the interaction 
between a BH binary and smaller objects early on in the history of the GC; the occurrence of Kozai resonance in triple systems, 
which can lead to mergers which otherwise would not take place within the Hubble timescale; and the tidal capture of stars by a massive BH in the center of the GC.

\subsubsection{Two-body encounters}

We start with the simplest to calculate, which is the rate of direct captures, i.e.\ two BHs that undergo a close encounter such that 
the energy loss to GWs is enough to become bound. The cross section for such an interaction has been calculated in
Refs.~\cite{QuinlanShapiro1989, MouriTaniguchi2002} and is given by
\begin{eqnarray}
\sigma_{\textrm{DC}} (m_{1}, m_{2}, v) &=& 2 \pi \left( \frac{85 \pi}{6\sqrt{2}} \right)^{2/7}  \left( \frac{v_{\rm rel}}{c}\right)^{-18/7} \\ 
&\times&  \left( \frac{G^{2} m_{1}^{2/7} m_{2}^{2/7} (m_{1}+m_{2})^{10/7}}{c^{4}}\right),  \nonumber
\label{eq:DirecrCaptCS}
\end{eqnarray}
where $m_{1}$ and $m_{2}$ are the two BH masses and $v_{\rm rel}$ their relative velocity. The direct capture rate is then just
\begin{equation}
R_{\textrm{DC}} = 4 \pi \int_{0}^{r_\textrm{max}} dr \, r^{2} \frac{1}{2} n_{\textrm{BH}}(r)^{2} \sigma_{\textrm{DC}}(m_{1}, m_{2}, v_{\textrm{rel}})v_{\textrm{rel}},
\label{eq:DirecrCaptRate}
\end{equation}
where $v_{\textrm{rel}} = \sqrt{2}\, v_{\textrm{DM}}$ and $n_{\textrm{BH}}$ the BH number density. When these events take place, 
the created binaries typically have very high initial eccentricities and small pericenter distances, and thus the merger
timescale in these systems is very short \cite{PetersMathews1963, Peters1964}, typically $10^{2}-10^{4}\,{\rm yrs}$ \cite{Cholis:2016kqi}.

Without mass segregation, the typical separation distance of  BHs in the core would be $\sim 10^{5}\,{\rm au}$, 
even if all BHs from the original stars remained in the cluster. This would lead to a depressing seed-BH merger-rate range of 
$R_{\textrm{DC}} \simeq 6 \times 10^{-12}\,{\rm yr^{-1}}$ in a GC like NGC 104 to $9 \times 10^{-16}\,{\rm yr^{-1}}$ in a GC 
similar to NGC 5927. However, as we demonstrated earlier, mass segregation happens very early, especially for the more massive objects. 
For BHs the radius that they are enclosed to within the cluster is $\sim 1-10\,{\rm pc}$ \cite{Morscher:2014doa, 2016MNRAS.463.2109R}. 
Restricting ourselves only to direct captures happening within the inner $r<1\,{\rm pc}$, and taking a realistic fraction $f^{\textrm{BH}}=0.02-0.2$ of the created GC BHs to remain within that radius, we get a seed-BH merger rate of $R_{\textrm{DC}} \simeq 2 \times 10^{-10}
 (f^{\textrm{BH}}/0.2)^{2}\,{\rm yr^{-1}}$ for NGC 104 and $2 \times 10^{-14} (f^{\textrm{BH}}/0.2)^{2}\,{\rm yr^{-1}}$ for 
 NGC 5927. These are about two orders of magnitude higher than the rates without mass segregation.  The 
 more massive GCs, such as NGC 104, become the dominant merger sites since they retain a larger faction of BHs 
 and had many more massive stars to begin with. However, with typically $O(1)$ mergers per Hubble time, even these massive GCs are unlikely to support a runaway formation of IMBHs resulting  from direct two-body capture. 

\subsubsection{Three-body hardening}

The next relevant BH-merger scenario is that in which an existing BH binary in the cluster hardens via three-body interactions and then
merges. To calculate the rate for this process, we first need to determine if there is a density cusp towards the center of the GC in the 
presence of an early-formed massive BH. Such a BH would have a radius of influence $R_{\textrm{in}} 
= G m/v_{\textrm{DM}}^{2}$, which is $5 \times 10^{-4}\,{\rm au}$  for a $30\,M_{\odot}$ in a GC like NGC 104, and $10^{-2}\,{\rm au}$ in 
a GC like NGC 5927. The number of segregated seed BHs in the inner $1\,{\rm pc}$ that will fall inside the radius of influence 
of the early massive BH is given by \cite{mandel08}
\begin{equation}
N_{\textrm{in}} = \frac{16 \pi}{5} R_{\textrm{in}}^{3} n_{\textrm{BH}},
\end{equation}
which amounts to only $0.3\,(7) \times 10^{-3}$, even for the most optimistic smaller GCs. Thus no cusp is 
formed in these stages. At later stages when the IMBH has grown to $O(10^{3})\,M_{\odot}$, a cusp can potentially be formed, 
but this is not relevant for the conditions that this work is focused on. This simplifies the three-body interactions to the case 
of isolated binary-single interactions. In that regime, there are two timescales that are important. The first is the timescale 
for a binary made up of a massive BH and another (seed) BH  to interact with third lighter bodies in the GC. 
The second is the merger timescale due to GW emission for a binary system with an eccentricity $e$ and a 
semi-major axis $a$. Note that since this binary is made up of a massive BH and a seed BH, substitution of the lighter companion may
take place in these interactions, occasionally resulting in the former companion getting ejected from the cluster \cite{Haster:2016ewz}. 
Ref.~\cite{mandel08} finds that in a $100\, M_{\odot}-10\, M_{\odot}$ system, only one per $5 \times10^{4}$ encounters with 
third objects leads to the $10\,M_{\odot}$ BH getting swapped. The typical timescale 
for that is well above the Hubble timescale. For simplicity we ignore the subtitutions of BHs via many-body interactions altogether.

Following the assumptions by \cite{mandel08} we take as typical values $a \sim10^{13}$ cm and $e=0.98$ for a binary 
with $m_{1} \gg m_{2}$. The merger timescale from GW emission is,
\begin{equation}
\tau_{\textrm{merge}} = 10^{8} \frac{1}{m_{2}} \frac{100}{m_{1}} \, \textrm{yr},
\end{equation} 
while the timescale for hardening by interaction with other bodies is,
\begin{equation}
\tau_{\textrm{harden}} =  \frac{2 \pi}{22} \frac{m_{1}+m_{2}}{m_{s}} \frac{1}{\dot{N}}\, \textrm{yr}.
\end{equation}  
Here $m_{s}\sim 0.5 M_{\odot}$ is the mass of a star and $\dot{N}=n v_{DM} \pi a (2GM/v_{\rm DM}^2)$ is the rate of 
interactions of the single bodies with the binary \cite{mandel08}.
The resulting merger rate from the three body interactions is simply $1/(\tau_{\textrm{merge}}  + \tau_{\textrm{harden}})$.
This should be considered as a conservative estimate, as in practice even higher eccentricities than $e=0.98$ can be reached during
the sequence of three-body interactions, possibly reducing the merger timescale \cite{Antonini:2013tea,Samsing:2013kua}.
For NGC 104, 6362 and 5927 we get $10^{-9}\,\textrm{yr}^{-1}$, $10^{-11}\,\textrm{yr}^{-1}$ and $7\times10^{-12}\,\textrm{yr}^{-1}$, respectively, depending weakly on the exact value of $m_{1}$. In this case both merger trees of Fig.~\ref{fig:IMBHGrowth} are competitive and viable at the early stages of the GCs. As in the direct capture case, the rate is again dominated by the massive end of the GC population.  Note that we have assumed here that stars have not been entirely kicked out of the inner $1\,{\rm pc}$, which may not always be true for older systems, but for GCs in their earlier stages it will still be relevant. Lighter objects will move outwards, but using Eq.~(\ref{eq:tauRelax}) and
Eq.~(\ref{eq:tauDF}) one can see that in the first $\sim\,{\rm Gyr}$ of the GC lifetime, enough of them will be still in the core of the 
cluster. If we consider that only seed BHs remain in the inner $1\,{\rm pc}$, then the hardening timescale of the initial binary 
from three-body interactions becomes very large (and inefficient), especially since the number density and as a result $\dot{N}$ is very small. 
  
\subsubsection{Other mechanisms}

The two-body capture and the three body effects between binary BHs and third objects are the two 
most straightforward effects to include. The uncertainties regarding the conditions in the core of GCs leading to the 
formation and growth of an IMBH in the center of GCs, while large, can be included in the calculations. Yet, there are 
other effects to discuss. These include the hardening of existing binaries via Kozai-Lidov resonance in triple systems and the capture of regular stars and of Neutron Stars (NS) from BHs via tidal effects. 

The Kozai resonance \cite{Kozai:1962zz} is relevant here when a third body, or potentially a 
second wider binary, excites an already tight binary into a highly eccentric orbit. During that time the binary loses energy 
in more modes and coalesces faster \cite{Miller:2002pg}. Ref.~\cite{OLeary:2005vqo} has shown that a second binary 
with the appropriate distance, masses and orbital properties to affect the merger timescale of the tight binary is rare 
and thus we ignore it. For a three-body system, where the tight binary, composed of seed BHs has a semi-major axis of $a \sim 1\,{\rm au}$, and the third stellar body of mass $\sim M_{\odot}$ has a semi-major axis of $O(10)\,{\rm au}$, 
the Kozai timescale becomes $\sim 10^{2}-10^{3}\,{\rm yrs}$. This is long enough to cause faster emission of GWs 
and harden the binaries. Yet, this effect is mostly relevant in the first Gyrs of the GC lifetime when such systems are 
going to be more common.  If the third object is a seed BH, the Kozai timescale is reduced by an order of magnitude, 
making its effect on the merger of the tight binary insignificant. We thus consider the occurrence of Kozai resonance to act as a possible enhancement of the merger rates in GCs at the early stages, providing further motivation for the runaway growth of IMBHs in GCs. 
  
 A massive BH at the center of a GC can also grow from the tidal capture of a main sequence star, resulting in
 infall of matter. Some of these systems can become Ultra-Luminous X-ray (ULX) sources. There is approximately 
 one ULX source per galaxy \cite{Hopman:2005hm}, or $6 \times 10^{-3}$ per GC. About $1\%$ of these will merge,
producing GWs that could show up in the LIGO band \cite{MacLeod:2015bpa}, but with a distinctively different waveform signal. Nevertheless, 
as many more stars around the  BH will experience partial accretion of their mass we cannot exclude this channel as 
a contributor to the overall  growth of the IMBH. 
Finally, Ref.~\cite{mandel08} has calculated the tidal effects that an existing massive BH has on a NS already spiraling 
around it, and found the excitation of NS modes to be too weak to affect the NS merger rate. 
We do not discuss the impact that NSs in GCs can have on the growth of a IMBH. Given the Kroupa mass function 
\cite{Kroupa:2000iv}, NSs could possibly contribute to the total IMBH mass only up to as much mass as the seed BHs. As NSs will segregate slower than BHs and into a larger volume being further separated from the center of the GC, we 
expect their contribution to the early growth of an IMBH to be small compared to the growth from BH merger events. 

\subsection{Binary ejection}

A crucial element in the scenario where IMBHs form by a runaway process of seed BH mergers is that enough of the post-merger binaries
are retained within the cluster after the merger to allow the process to continue. Past simulations have cast doubt on this, as they repeatedly
showed that most binaries get ejected from the cluster due to three-body interactions at some point during the hardening process \cite{PortegiesZwart:1999nm,Rodriguez:2015oxa}. They can 
also be ejected to gravitational-wave recoil \cite{Miller:2002vg,Haster:2016ewz,HolleyBockelmann:2007eh}. This picture, however, was based on calculations that did not take into account post-Newtonian 
corrections leading and up to 2.5pN order \cite{Haster:2016ewz}. As mentioned in the Introduction, recently, both analytic \cite{Samsing:2017xmd} and numerical \cite{Gondan:2017wzd} analyses incorporating these effects have shown that orders-of-magnitude more binaries are retained in the GC after merger, rekindling the runaway merger of BHs as a viable scenario for IMBH formation. A precise treatment of this issue is well beyond the scope of this work, but as these new calculations show that roughly half of the merging binaries are still ejected from the cluster prior to the merger, the more massive GCs (such as NGC 104), which have more BHs to begin with, should be considered the preferred sites for this process to occur.  

\subsection{Total occupation fraction}

To sum up this Section, we return to the question of what our expectation should be for $f_{\rm occ}$, the overall occupation fraction of runaway-formed IMBHs in GCs. 
From our results, it is likely that only the most massive and dense systems, such as NGC 104, can allow for the required 
merger and retention rates to support a runaway merger leading to the formation of an IMBH. Examining the list of 157 observed MW clusters, 
we find that approximately $\sim10\%$ have similar properties (density, mass, velocity dispersion) to NGC 104. 
 Assuming that the observed sample of MW GCs is typical for GCs in all galaxies and based on the rates estimated above, 
 we therefore conservatively conclude that $f_{\rm occ}$ is unlikely to exceed $10\%$ by much. This sets a clear target for aLIGO to probe as deep as 
 possible into the regime $f_{\rm occ}<10\%$. Fortunately, the results of this work will show that this regime is penetrable with 
 aLIGO running at design sensitivity for a period of $O(5)$ years!


\section{Background}
\label{sec:Background}

Inevitably, the mergers we are after will have to be distinguished from the background events that have
nothing to do with IMBH formation.  Referring again to Eq.~(\ref{eq:Nobs}), we require a prescription for the mass
function and rate of background merger events in order to calculate $N^{\rm BG}_{\rm obs}(M)$.
To quantify the mass distribution, we employ 
a simple ansatz for the mass function of a merging BH binary with component masses $M_1>M_2$ 
(see \cite{Kovetz:2016kpi,Kovetz:2017rvv})
\begin{equation}
\frac{dN(M_1)}{dM_1} = P({M_1}) \int\limits_{M_{\rm gap}}^{M_1} P(M_2)dM_2,
\label{eq:BHMF}
\end{equation}
where
\begin{eqnarray}
P(M_1)&=& A_{M_1}M_1^{-\alpha}\mathcal{H}(M_1\!-\!M_{\rm gap})e^{-(M_1/M_{\rm cap})^2}, \nonumber \\
P(M_2)&=& A_{M_2}(M_2/M_1)^{\beta}\mathcal{H}(M_2\!-\!M_{\rm gap})\mathcal{H}(M_1\!-\!M_2). \nonumber \\
\end{eqnarray}
Here $\alpha$ is a power law with a fiducial value of $2.35$ (to match the Kroupa mass function \cite{Kroupa:2000iv}), 
$M_{\rm gap}$ is the lowest stellar-BH mass possible, which we set to $5\,M_\odot$ (this has no effect on our results),  
$M_{\rm cap}$ is a double-exponential upper cutoff on the stellar-BH mass, $\beta$ is the power-law index of the mass ratio, whose choice generally depends on the binary-BH progenitor model, $\mathcal{H}$ is the Heaviside function and $A_{M_1}, A_{M_2}$ are normalization constants. As default values we take throughout $M_{\rm cap}=40\,M_\odot$ and $\beta=0$, both consistent with current aLIGO observations.

For the rate of mergers, we take the central value from Ref.~\cite{Abbott:2017vtc}, $\bar{R}_{\rm BG}=103\,{\rm Gpc^{-3}~yr^{-1}}$.
As we did when calculating the signal, in the simplest case we will take the rate to be redshift independent. To account for time evolution, 
we will correct the background rate at $z=0$ according to \cite{TheLIGOScientific:2016wyq,Cholis:2016xvo,Abbott:2017xzg}
\begin{equation}
\bar{R}_{\rm BG}(z)=\int\limits^{t_{\rm max}}_{t_{\rm min}}R_f(z_f)P(t_d)\,dt_d,
\label{eq:RzTd}
\end{equation}
where $z_f$ is the redshift at formation, $R_f(z_f)$ is the best-fit function found in Ref.~\cite{Madau:2014bja} for the star-formation-rate history and 
the convolution is with a delay-time distribution of the form $P(t_d)=1/t_d$, with $t_{\rm min}=50\,{\rm Myrs}$ and $t_{\rm max}$ equal
to the Hubble time.  

Our fiducial model includes a double-exponential cutoff at high mass---supported both by current data from aLIGO and 
by theoretical models of pair-instabillity (and pulsational pair-instability) SNe which predict a dearth of stellar BHs above $M\sim50\,M_\odot$ \cite{Spera:2016slz,Spera:2017fyx})---and $\beta=0$, so that 
$M_2$ takes values uniformly from $M_{\rm gap}$ to $M_1$. However, to account for the uncertainty regarding these 
choices, we will calculate our results for a variety of different models, including a shallower exponential cutoff as well as a sharper cutoff, different values for $\beta$, and a value of $R_{\rm BG}$ corresponding to the upper bound at $90\%$ confidence in the recent aLIGO analysis \cite{Abbott:2017vtc}.


\section{Advanced LIGO Sensitivity}
\label{sec:LIGOSensitivity}

Our observable, introduced in Eq.~(\ref{eq:Nobs}) in Section~\ref{sec:setup}, depends on the quantity $\langle VT\rangle$,
the detectable space-time volume of aLIGO, which in turn is given by \cite{Abbott:2017iws} 
\begin{equation}
\langle VT\rangle=T_{\rm obs}\int dzd\theta \frac{dV_c}{dz}\frac{1}{1+z}s(\theta)f(z,\theta),
\label{eq:VT}
\end{equation}
where $T_0$ is the coincident online time of the LIGO interferometers, $V_c(z)$ is the comoving volume 
in a sphere that extends to redshift $z$, $s(\theta)$ is an injected distribution of binary parameters which include masses,
spins, tilt angles and the orientation of the orbital plane, and $f(z, \theta)$ is  the fraction of injections with redshift $z$ and parameters $\theta$ that are 
detectable by aLIGO with a given sensitivity (the standard criterion for detection is a signal-to-noise threshold of $8$ per detector). 
To evaluate the integral in Eq.~(\ref{eq:VT}) we follow Ref.~\cite{Chen:2017wpg} and use the IMRPhenomD waveform approximant given in the LALSimulation package  \cite{Veitch:2014wba,Smith:2016qas} to perform a Monte Carlo simulation. To get the noise power spectra for the different versions of aLIGO we consider, we digitized the curves in Ref.~\cite{Aasi:2013wya}. 
As a consistency check, we compared our results with the online GW Distance Calculator \cite{GWDistCalcwebsite} and verified that our calculations are consistent to within less than $\sim10\%$. 

\begin{figure}[h!]
\begin{centering}
\includegraphics[width=\columnwidth]{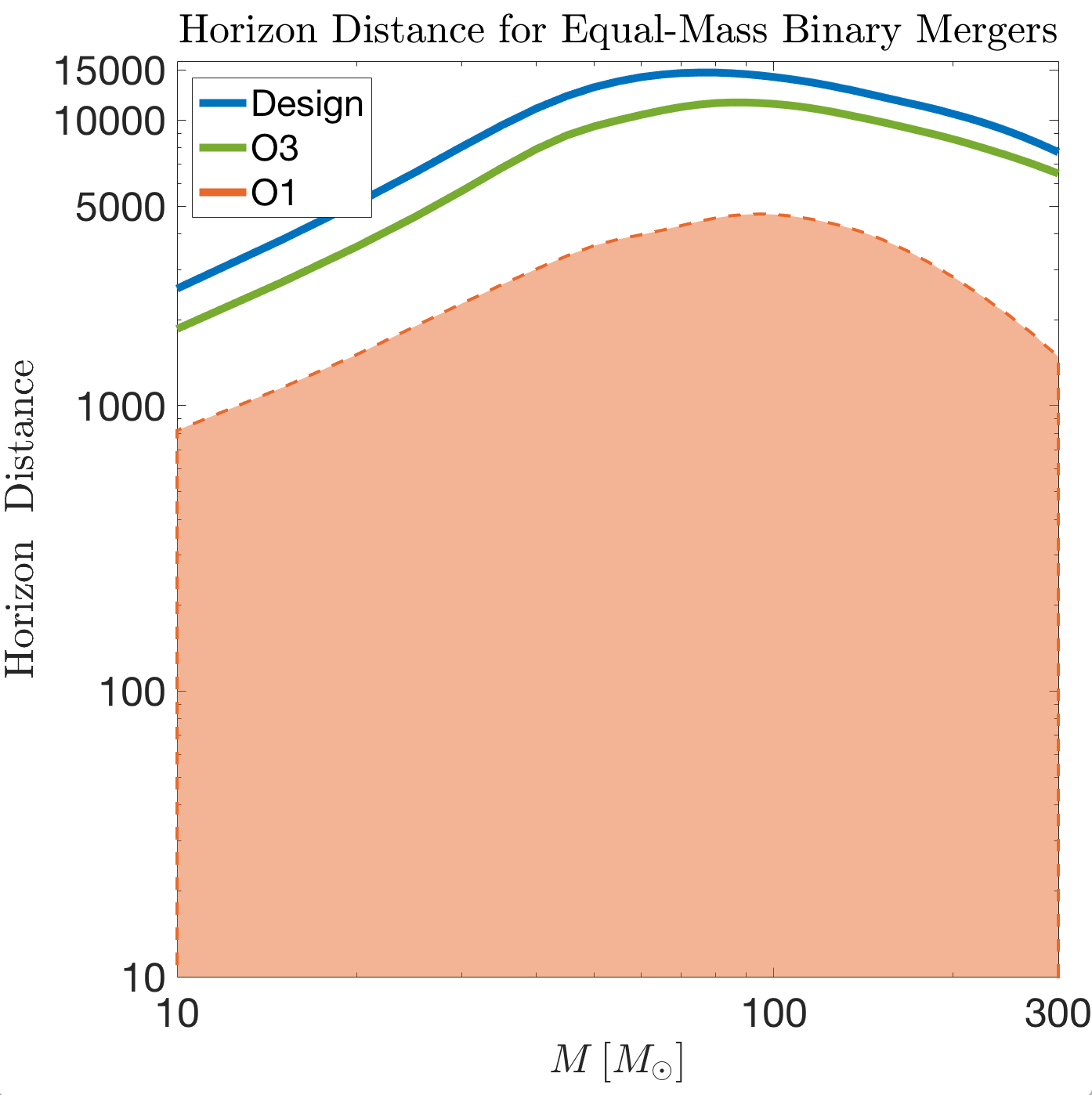}
\includegraphics[width=\columnwidth]{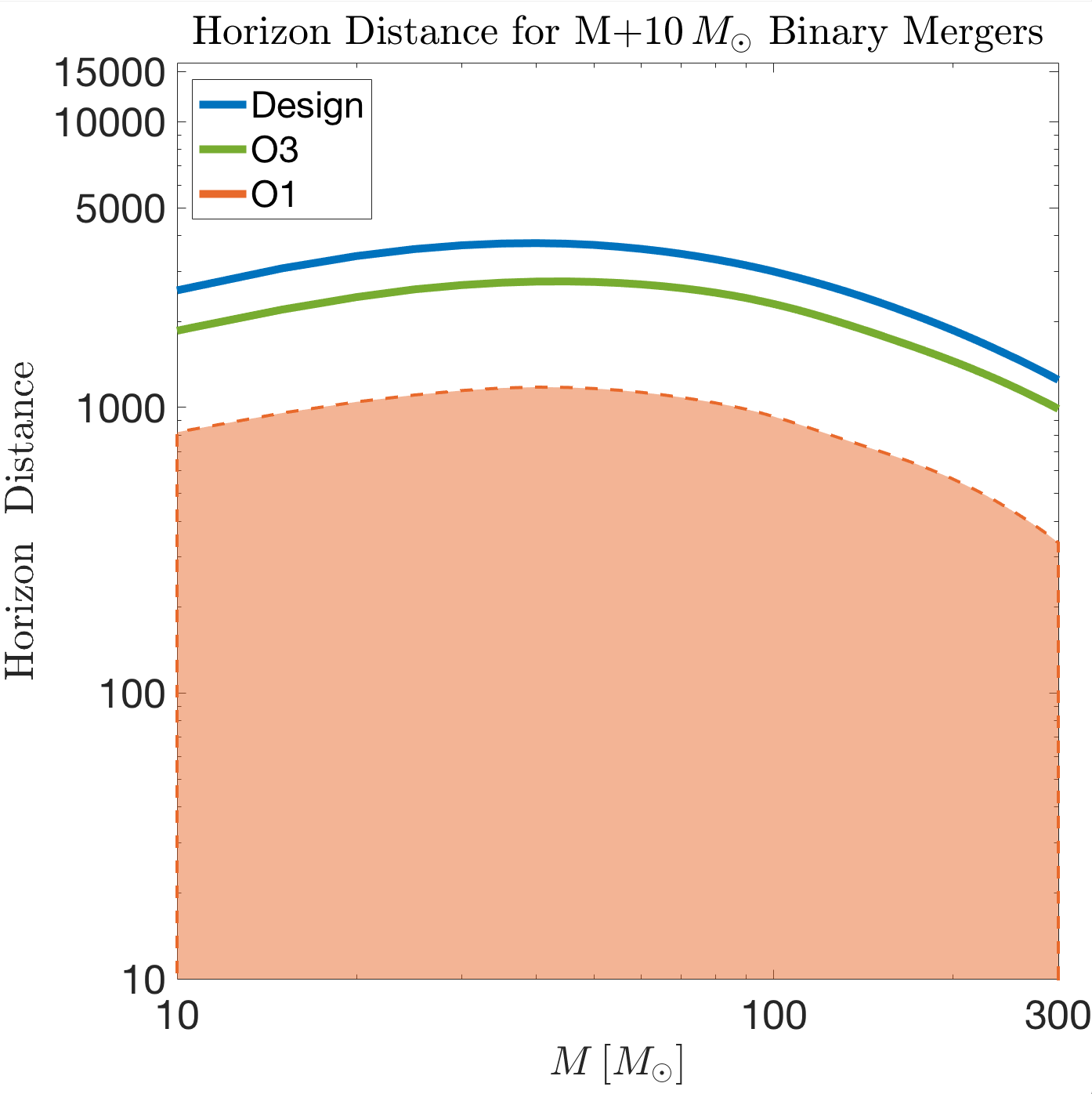}
\end{centering}
\caption{The sensitivity of different versions of aLIGO to mergers of equal-mass binaries ({\it top}) and binaries with a massive BH and a $10\,M_\odot$ seed BH ({\it bottom}). In both cases, the sensitivity is shown as a function of the mass of the heavier BH.}
\label{fig:LIGOSensitivity}
\end{figure}

In Fig.~\ref{fig:LIGOSensitivity} we plot the horizon distance---the farthest luminosity distance at which a source could ever be detected---for mergers involving equal-mass binaries, matching our bottom-heavy Isosceles Triangle merger tree, as well as for ones that involve a massive BH merging with a seed BH, corresponding to the top-heavy Right Triangle merger tree. We see that for equal-mass binaries, the sensitivity peaks when the two BHs are roughly $100\,M_\odot$, while in the Right Triangle case the sensitivity peaks earlier and then slowly decreases. It is important to note that in the latter case, when the massive BH approaches $300\,M_\odot$, the mass ratio starts to become large, and the waveforms for the signal from these mergers should be considered less reliable \cite{Smith:2013mfa}.
  
\begin{figure}[h!]
\begin{centering}
\includegraphics[width=\columnwidth]{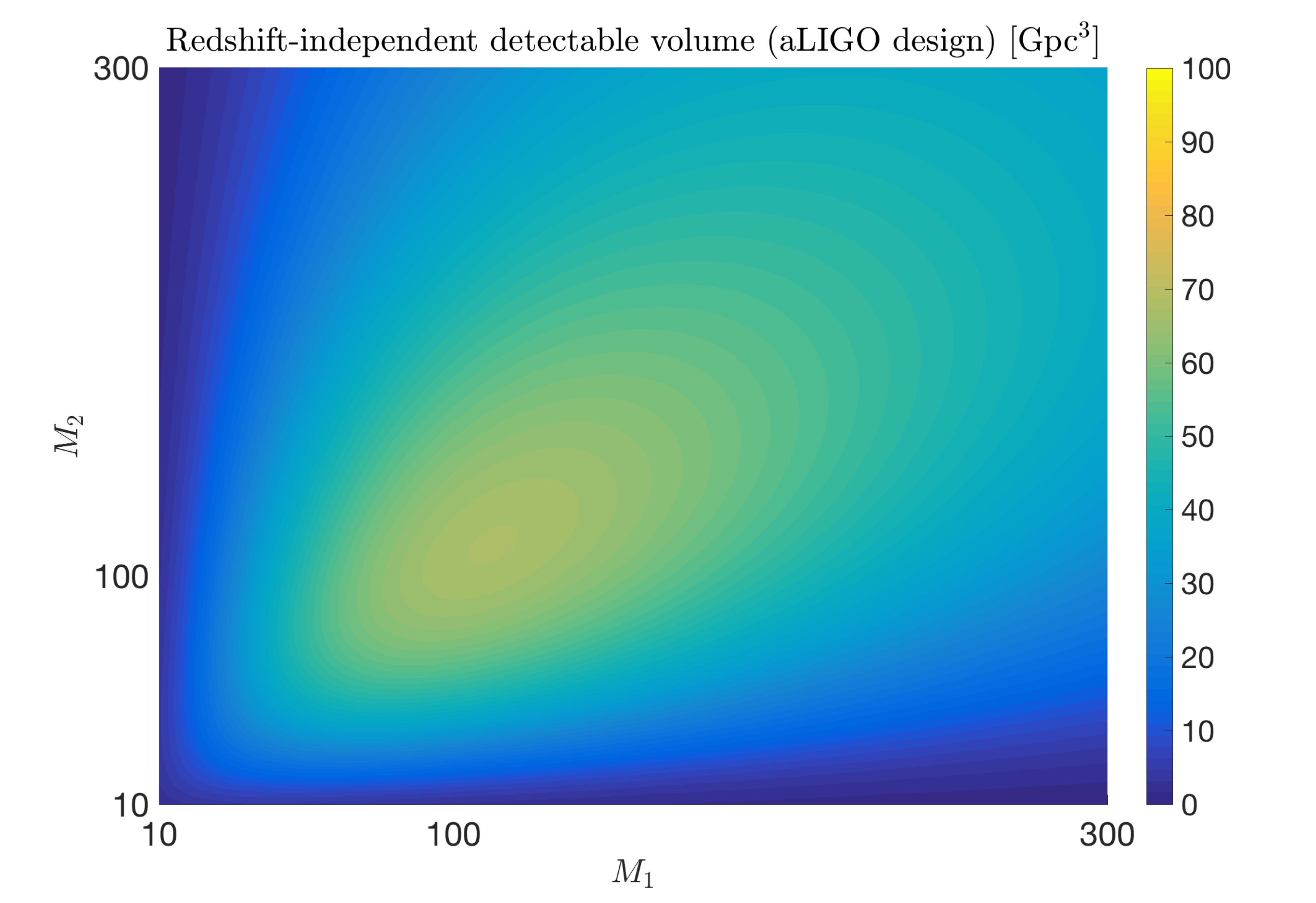}
\includegraphics[width=\columnwidth]{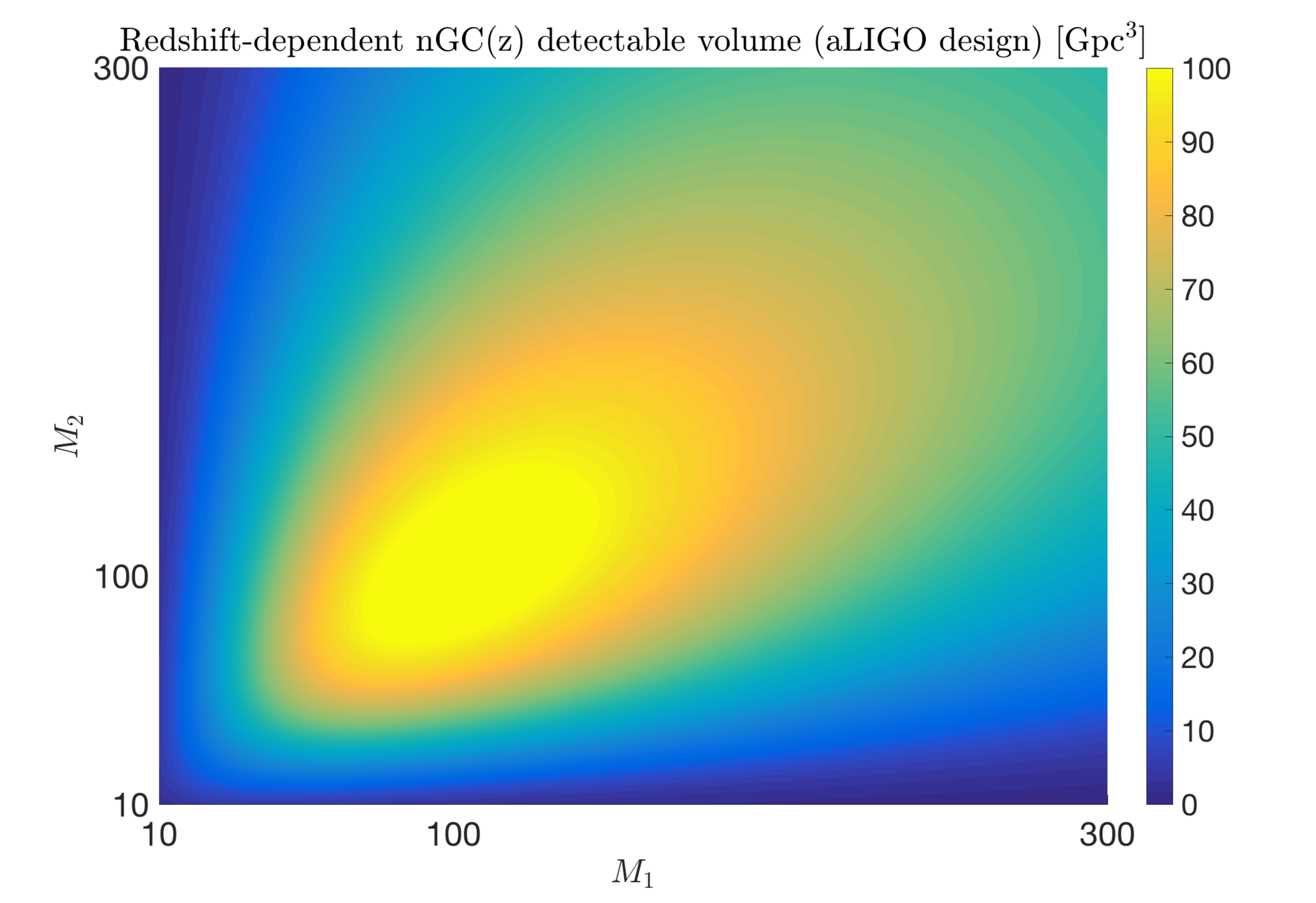}
\includegraphics[width=\columnwidth]{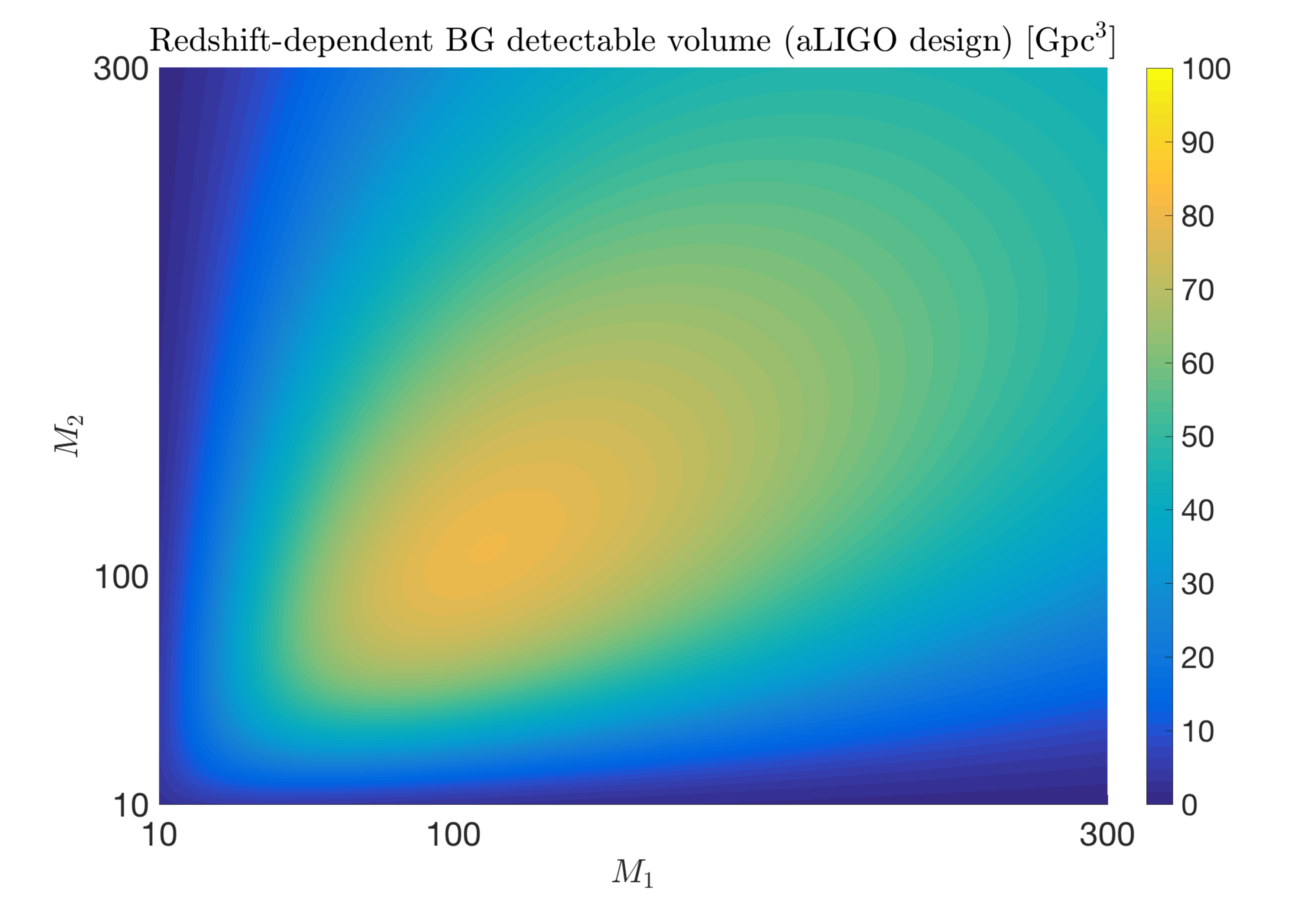}
\end{centering}
\caption{The detectable volume of aLIGO at design sensitivity for a redshift-independent merger rate ({\it top}), compared with the detectable volume when incorporating the redshift dependence in our signal ({\it middle}) and background ({\it bottom}).}
\label{fig:LIGOVolSensitivity}
\end{figure}

To incorporate a redshift dependence for the signal and background, we correct the detectable volume calculated via Eq.~(\ref{eq:VT}) by 
a factor $\bar{V_z}/\bar{V_0}$.  Here, $\bar{V_0}$ is the average detectable volume for all sources, calculated by weighting the comoving volume at redshift $z$ by the unit detectable volume at each redshift slice, and $\bar{V_z}$ is a similar average volume with the weights multiplied by the appropriate redshift-dependence for either the signal (i.e.\ $n_{\rm GC}(z)$) or the background (i.e.\ Eq.~(\ref{eq:RzTd}), which accounts for a time-delayed star-formation-rate history); see Fig.~\ref{fig:LIGOVolSensitivity}.  
 
Lastly, we wish to conservatively take into account the mass-measurement uncertainty in the
observed mass distribution of detected events, especially at the high-mass end \cite{Haster:2015cnn}. To do so, we follow Ref.~\cite{Kovetz:2016kpi} 
and convolve the mass functions for the signal and background with a log-normal 
distribution
\begin{eqnarray}
P(M_{\rm obs})&=&\iint P(M_{\it th})P_{\rm G}(x)\delta\left(M_{\rm obs}-
xM_{\rm th}\right)dx\,dM_{\rm th}  \cr
&=&\int P(M_{\rm th})P_{\rm G}\left(M_{\rm obs}/M_{\rm th}\right)
dM_{\rm th}/M_{\rm th}.
\label{eq:Pmobs}
\end{eqnarray} 
Here, $M_{\rm th}$ is the true value of the mass (which follows the theoretical 
mass functions given in Eqs.~(\ref{eq:MFBottom}),~(\ref{eq:MFTop}) and (\ref{eq:BHMF})), 
$M_{\rm obs}$ is the observed mass, and the relation between them is given by 
$M_{\rm obs}=xM_{\rm th}$, where $x$ follows a normal distribution, $x\sim\mathcal{N}(1,\sigma^2)$ and 
$P_{\rm G}=\frac{1}{\sqrt{2\pi\sigma^2}}e^{-(x-1)^2/2\sigma^2}$.
We set the relative measurement error to  $\sigma=0.1$ ($10\%$). 


 \section{Results}
\label{sec:results}

With prescriptions in hand for calculating the signal and background, as well as taking into account the experimental sensitivity and measurement uncertainty properly, we are ready to present our results. We start from the observed mass distribution, Eq.~(\ref{eq:Nobs}), and then proceed to calculate the limits on $f_{\rm occ}$ that aLIGO will be able to set in its future runs (as discussed below, we find that the O1 and O2 runs that have recently been completed do not yet place meaningful constraints on the runaway formation of IMBHs in GCs). 

\subsection{The observed mass distribution with aLIGO}

\begin{figure}[h!]
\begin{centering}
\includegraphics[width=\columnwidth]{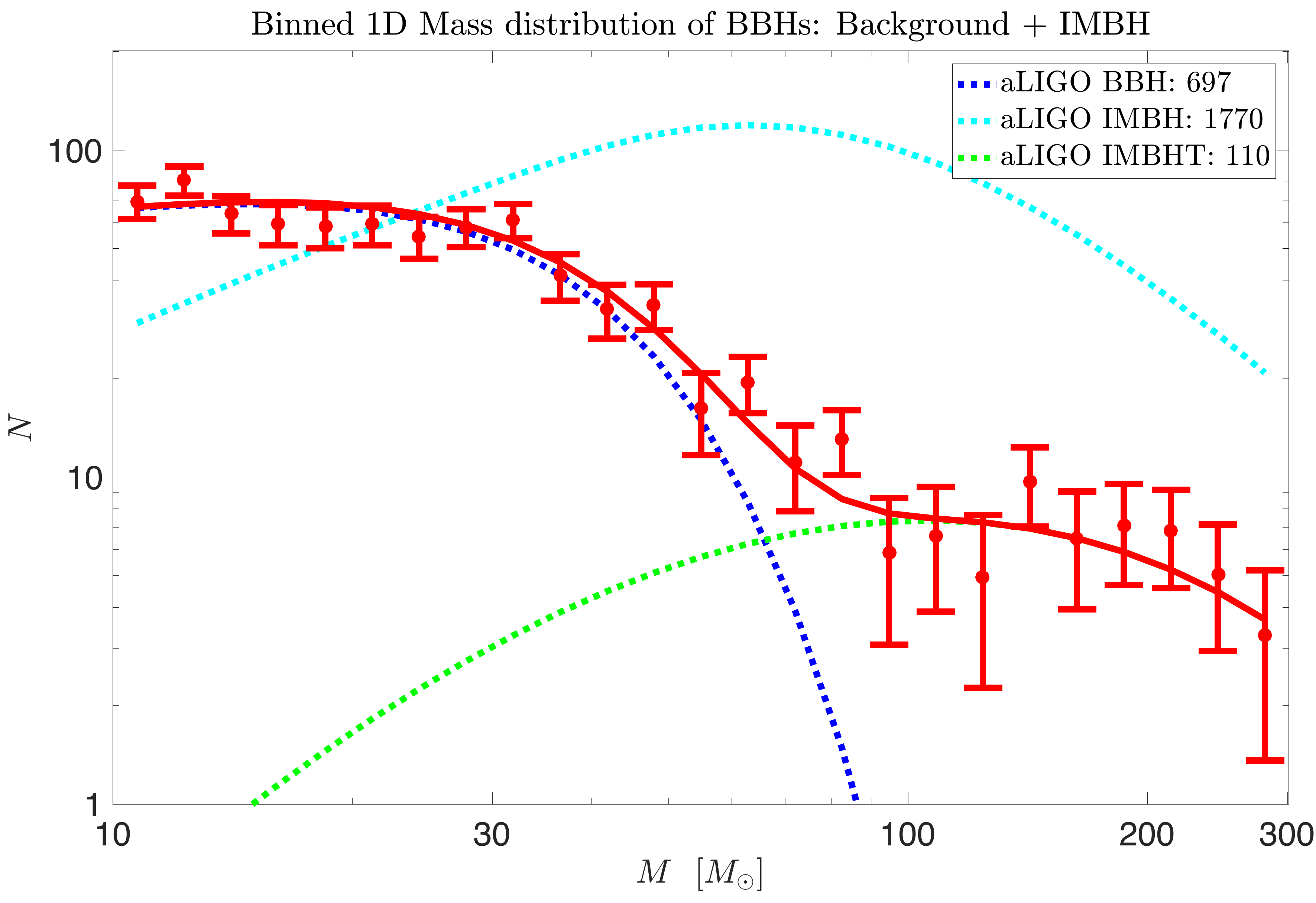}
\includegraphics[width=\columnwidth]{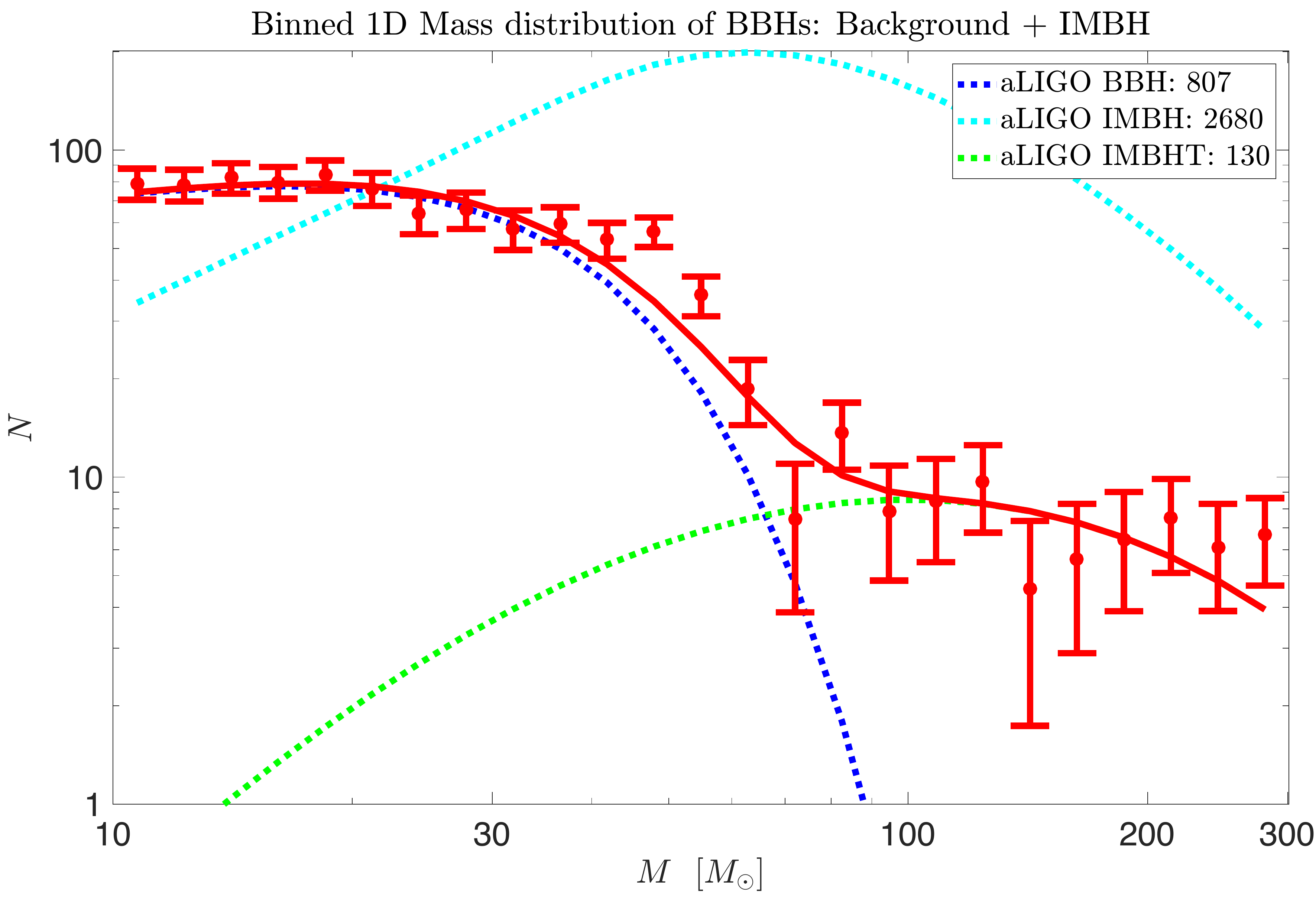}
\end{centering}
\caption{The observed mass distribution of GW mergers detectable by aLIGO running for $6$ years at design sensitivity. The dashed blue curve shows the distribution for the background model with a double-exponential cutoff at $M_{\rm cap}=40\,M_\odot$. The dashed-cyan (dashed-green) show the predicted mass distribution from mergers leading to IMBH formation via the Isosceles (Right) Triangle merger tree described in Section~\ref{sec:Signal}. The red curve shows the total number of observed BHs, with Poisson errors in each logarithmic bin. 
{\it Top}: No redshift dependence. {\it Bottom}: taking into account the redshift evolution of the merger rate, using $n_{\rm GC}(z)$ for the signal as described in Section~\ref{sec:Signal} and the time-delayed star-formation-rate history for the background (see Section~\ref{sec:Background}).}
\label{fig:LIGOMassDistribution}
\end{figure}

In Fig.~\ref{fig:LIGOMassDistribution} we show our prediction for the number of observed mergers as a function of the heavier BH mass in the merging binary, for aLIGO running at design sensitivity. We calculate this twice, with and without taking into account the redshift evolution of the merger rate, as explained in the previous sections. 

We chose to calculate $N_{\rm obs}(M)$ up to $M=300\,M_\odot$, both to avoid using waveforms for mass ratios larger than $30$ and to conservatively focus on the lighter IMBHs in the range we are interested in. We note that if the runaway formation process tends to be more effective and generally yields even more massive IMBHs, our limits on $f_{\rm occ}$ will be a safe underestimate. For consistency we therefore set $M_{\rm IMBH}=600\,M_\odot$ when calculating the mass function for the Isosceles scenario (as that is the lightest IMBH that can be formed from a merger of $300\,M_\odot$ equal-mass BHs). The background parameters used here are: a double-exponential cutoff with $M_{\rm cap}=40$, $\beta=0$ and $\bar{R}_{\rm BG}=103\,{\rm Gpc^{-3}yr^{-1}}$. The binning is logarithmic with roughly $10\%$ mass width. 
Evidently, even with the minimal rate (for the Right Triangle case), a null detection of mergers at high mass will rule out $f_{\rm occ}=100\%$.  
We infer a more precise limit below.

\subsection{Limits on runaway IMBH formation}

In a Poisson distribution, the probability to observe $n_i$ events in a mass bin $i$ given an expected value $\mu_i$ is $P(n_i,\mu_i)= \mu_i^{n_i}e^{-\mu_i}/n_i$. Therefore, the log-likelihood for observing a set $n_i$ of events is given by
\begin{equation}
\ln L = \sum\limits_i \ln\left(\frac{\mu_i^{n_i}e^{-\mu_i}}{n_i!}\right).
\end{equation} 
In order to derive our upper bound on $f_{\rm occ}$, we generate an ensemble of $100,000$ expected log-likelihoods where the $n_i$ are randomly drawn from a Poisson distribution characterized by $\mu_i=N^{\rm BG}_{i}$, to reflect the null hypothesis of no signal from IMBH formation, and then solve for the value of $f_{\rm occ}$ for which the median of another sample of $500$ log-likelihoods, this time drawn from a Poisson distribution with $\mu_i=N^{\rm Signal}_{i}+N^{\rm BG}_{i}$ is larger than a fraction $0.001$ of the original ensemble. This  yields our $99.9\%$-confidence upper bound on $f_{\rm occ}$.

\begin{table*}
\def\arraystretch{1.75}
    \begin{tabular}{|cc|c|c|c|c|c|c|c|c|}
         \hline
   \multicolumn{2}{|c|}{}      &\multicolumn{4}{c|}{Right Triangle}  & \multicolumn{4}{c|}{Isosceles} \\
\hline      
\multicolumn{2}{|c|}{Background Model} & O1+O2& O3 (1 yr)& Design (6 yrs)& with $R(z)$ &  O1+O2& O3 (1 yr)& Design (6 yrs) & with $R(z)$\\
            \hline
\multicolumn{2}{|c|}{~$P(M)\propto\exp({-(M/40)^2})$~}& $470\%~(0.2) $&$ 38\% ~(3)$& $3.1\% ~(47)$& $2.9\% ~(53)$& $19\% ~(3.9)$& $3.4\%~ (34)$ & $ 0.4\% ~(420)$ &$0.3\%~ (630)$ \\
    		 \multicolumn{2}{|c|}{~$P(M)\propto\exp({-(M/60)^2})$~}&  $760\%~(0.2) $ & $68\%~ (3)$ & $7.4\%~ (47)$& $ 6.8\% ~(53) $&$34\%~ (3.9)$& 6.3\% (34)&$1.2\%~(420)$&$ 0.3\%~ (630)$ \\
     		 \multicolumn{2}{|c|}{$P(M)\propto\mathcal{H}(50-M)$}& $220\%~ (0.2)$ & $17\%~ (3)$ & $1.3\%~ (47)$& $1.2\%~ (53)$&$9.7\% ~(3.9)$& 1.4\% (34)&$0.2\%~ (420)$&$0.08\%~ (630)$ \\
     		 \multicolumn{2}{|c|}{$P(M)\propto\exp({-M/40})$}&$ 890\%~(0.2) $& $120\% ~(3)$ &$ 19\%~ (47)$&18\% (53)&$36\%~ (3.9)$& $9.2\% ~(34)$ &$2.2\%~ (420)$& $1.7\% ~(630)$\\
     		 \multicolumn{2}{|c|}{$\bar{R}_{\rm BG}=103+110=213$}&$ 470\%~ (0.2)$& $41\% ~(3)$ &$ 3.4\%~ (47)$&$3.1\%~ (53)$ &$22\% ~ (3.9)$& $3.5\%~ (34)$ &$0.5\%~ (420)$&$0.4\%~ (630)$ \\
      		 \multicolumn{2}{|c|}{$\beta=-1$}& $ 430\% ~(0.2)$& $34\%~ (3)$ & $2.8\%~ (47)$& $2.7\%~ (53)$&$19\% ~(3.9)$& $2.9\% ~(34)$&$0.4\%~ (420)$&$0.3\%~ (630)$ \\
    		 \multicolumn{2}{|c|}{$\beta=1$}& $ 480\% ~(0.2)$ & $34\% ~(3)$ &$3.3\%~ (47)$&$3.1\%~ (53)$&$22\% ~ (3.9)$&$ 3.3\% ~(34)$ &$0.5\%~ (420)$& $0.4\%~ (630)$\\
            \hline             
        \end{tabular}
    \caption{$99.9\%$ confidence-level contraints on $f_{\rm occ}$, the occupation fraction of IMBHs formed in a runaway process (with the two merger trees considered above: Right triangle and Isolsceles) in globular clusters. The predicted number of IMBH-related observed events with $M>100\,M_\odot$ in each scenario (for $f_{\rm occ}=100\%$) is shown in parentheses. See main text for more details.}
    \label{tab:focc}
\end{table*}
 
We show our results in Table~\ref{tab:focc}. To make sure our derived limits are robust, we survey various possible models for the mass distribution of background mergers, from a mass function with a sharp cutoff at $50\,M_\odot$, to ones with a weak exponential cutoff at $40\,M_\odot$ or a double-exponential cutoff at either $40\,M_\odot$ (our default model) or $60\,M_\odot$. We also vary $\beta$ to reflect background models with tendency for equal-mass mergers (which may be the case for dynamically-formed binaries in GCs \cite{Park:2017zgj}) or for large mass ratios. Finally, we repeat the calculation for a background rate consistent with the $90\%$ upper bound on the merger rate given in Ref.~\cite{Abbott:2017vtc}. Conservatively focusing on the lower merger rate of the Right Triangle scenario, we find that current aLIGO measurements (from the O1 and O2 runs) do not place any constraints on the runaway formation of IMBHs in GCs. However, for all our models we find that aLIGO at design sensitivity will be able to probe well below $f_{\rm occ}=10\%$, with most cases yielding a limit around $f_{\rm occ}\lesssim3\%$. We also find that accounting for the redshift dependence does not affect our results considerably. 

Comparing with our expectation for the occupation fraction in Section~\ref{sec:Expectation}, we see that aLIGO will be able to produce valuable limits on the runaway scenario within the decade, shedding light on the question of IMBH formation in GCs directly via observation (or non-observation) of GWs from mergers of BHs that are accessible to aLIGO. 
We can also contrast our results for the predicted rate with existing aLIGO limits on IMBH mergers \cite{Abbott:2017iws}:
for $100-100\,M_\odot$ binaries, the LIGO collaboration has recently derived a bound on the merger rate of $<1\,{\rm Gpc^{-3}yr^{-1}}$ at $90\%$ confidence (which as they report, is equivalent to a rate-per-GC of $0.3\,{\rm Gyr^{-1}}$ if the number density of GCs is $\sim3\,{\rm Mpc}^{-3}$, which is the same value we adopt in this work). This is consistent with our findings which show that the more prolific Isosceles merger tree is already in tension with observations if $f_{\rm occ}=100\%$ (in this case we predict $3.9$ events involving masses $>100\,M_\odot$, as indicated in Table~\ref{tab:focc}).


\section{Conclusions}
\label{sec:Conclusions}

In this work we addressed a question involving IMBHs that as we demonstrated, can be 
effectively answered with upcoming measurements of advanced LIGO at its design sensitivity. The question is 
whether IMBHs can form in globular clusters via runaway merger of stellar seed BHs. Our results show that
aLIGO will be able to probe the occupation fraction $f_{\rm occ}$ of IMBHs formed in this way in GCs down to values
as low as $f_{\rm occ}\lesssim3\%$ (at $99.9\%$ significance), if no GW events are detected from mergers involving BHs with masses in the
range $\gtrsim50-300\,M_\odot$. This will place a robust bound on this scenario, which comes directly from 
observations and is independent of the numerous assumptions about the structure and dynamics of GCs that inevitably 
enter any attempt to calculate this quantity from first principles. The rather strict limit we forecast is especially interesting,
as analytical estimates we carry out above, using empirical measurements of a representative sample of Milky-Way GCs, 
show that $f_{\rm occ}$ can be as large as $10\%$ or more. 

There is growing evidence that IMBHs may be ubiquitous in dwarf galaxies and GCs. If indeed most GCs harbor an IMBH at their centers, 
the limit we show can be derived from aLIGO measurements would mean that alternative scenarios to the runaway merger of stellar BHs 
are required to explain how they formed (such as POP III stars, direct collapse of massive gas clouds or mergers of stars followed by direct collapse).
It is exciting that this conclusion can be achieved with an experiment that has almost no direct access to mergers that involve IMBHs themselves.  
Naturally, future ground-based gravitational-wave observatories such as the Einstein Telescope \cite{ET,Sathyaprakash:2012jk} and Cosmic Explorer \cite{Evans:2016mbw}, as well as the 
proposed space interferometers DECIGO \cite{Seto:2001qf} and LISA \cite{AmaroSeoane:2012je,2017arXiv170200786A}, will be able to achieve more direct constraints, 
however those will require decades of suspense. Meanwhile results from aLIGO are already pouring in.

\acknowledgements
We acknowledge the use of the python notebook (https://github.com/hsinyuc/distancetool) written by Hsin-Yu Chen, which we adapted and extended in order to perform the aLIGO sensitive-volume calculations made in this work. IC thanks the organizers of the GW and Cosmology workshop in DESY, Hamburg, Germany. 
This work was supported by NSF Grant No. 0244990, NASA NNX15AB18G and NNX17AK38G and the Simons Foundation.


\begin{thebibliography}{95}
\expandafter\ifx\csname natexlab\endcsname\relax\def\natexlab#1{#1}\fi
\expandafter\ifx\csname bibnamefont\endcsname\relax
  \def\bibnamefont#1{#1}\fi
\expandafter\ifx\csname bibfnamefont\endcsname\relax
  \def\bibfnamefont#1{#1}\fi
\expandafter\ifx\csname citenamefont\endcsname\relax
  \def\citenamefont#1{#1}\fi
\expandafter\ifx\csname url\endcsname\relax
  \def\url#1{\texttt{#1}}\fi
\expandafter\ifx\csname urlprefix\endcsname\relax\def\urlprefix{URL }\fi
\providecommand{\bibinfo}[2]{#2}
\providecommand{\eprint}[2][]{\url{#2}}
  
  \bibitem{Mortlock:2011va} 
  D.~J.~Mortlock {\it et al.},
  Nature {\bf 474}, 616 (2011)
  [arXiv:1106.6088 [astro-ph.CO]].
  
 \bibitem{Ghez:1998ph} 
  A.~M.~Ghez, B.~L.~Klein, M.~Morris and E.~E.~Becklin,
  Astrophys.\ J.\  {\bf 509}, 678 (1998)
  [astro-ph/9807210]. 

\bibitem{Ghez:2008ms} 
  A.~M.~Ghez {\it et al.},
  Astrophys.\ J.\  {\bf 689}, 1044 (2008)
  [arXiv:0808.2870 [astro-ph]].
  
  \bibitem[{\citenamefont{Motch et~al.}(1997)\citenamefont{Motch, Haberl,
  Dennerl, Pakull, and Janot-Pacheco}}]{Motch:1996gw}
\bibinfo{author}{\bibfnamefont{C.}~\bibnamefont{Motch}},
  \bibinfo{author}{\bibfnamefont{F.}~\bibnamefont{Haberl}},
  \bibinfo{author}{\bibfnamefont{K.}~\bibnamefont{Dennerl}},
  \bibinfo{author}{\bibfnamefont{M.}~\bibnamefont{Pakull}}, \bibnamefont{and}
  \bibinfo{author}{\bibfnamefont{E.}~\bibnamefont{Janot-Pacheco}},
  \bibinfo{journal}{Astron. Astrophys.} \textbf{\bibinfo{volume}{323}},
  \bibinfo{pages}{853} (\bibinfo{year}{1997}), \eprint{[astro-ph/9611122]}.

\bibitem[{\citenamefont{{in't Zand} et~al.}(2000)}]{in'tZand:2000zz}
\bibinfo{author}{\bibfnamefont{J.~J.~M.} \bibnamefont{{in't Zand}}}
  \bibnamefont{et~al.}, \bibinfo{journal}{\aap} \textbf{\bibinfo{volume}{357}},
  \bibinfo{pages}{520} (\bibinfo{year}{2000}), \eprint{[astro-ph/0001110]}.

\bibitem[{\citenamefont{Grimm et~al.}(2002)\citenamefont{Grimm, Gilfanov, and
  Sunyaev}}]{Grimm:2001vd}
\bibinfo{author}{\bibfnamefont{H.~J.} \bibnamefont{Grimm}},
  \bibinfo{author}{\bibfnamefont{M.}~\bibnamefont{Gilfanov}}, \bibnamefont{and}
  \bibinfo{author}{\bibfnamefont{R.}~\bibnamefont{Sunyaev}},
  \bibinfo{journal}{Astron. Astrophys.} \textbf{\bibinfo{volume}{391}},
  \bibinfo{pages}{923} (\bibinfo{year}{2002}), \eprint{[astro-ph/0109239]}.

\bibitem[{\citenamefont{Lutovinov et~al.}(2005)\citenamefont{Lutovinov,
  Revnivtsev, Gilfanov, Shtykovskiy, Molkov, and Sunyaev}}]{Lutovinov:2004wi}
\bibinfo{author}{\bibfnamefont{A.}~\bibnamefont{Lutovinov}},
  \bibinfo{author}{\bibfnamefont{M.}~\bibnamefont{Revnivtsev}},
  \bibinfo{author}{\bibfnamefont{M.}~\bibnamefont{Gilfanov}},
  \bibinfo{author}{\bibfnamefont{P.}~\bibnamefont{Shtykovskiy}},
  \bibinfo{author}{\bibfnamefont{S.}~\bibnamefont{Molkov}}, \bibnamefont{and}
  \bibinfo{author}{\bibfnamefont{R.}~\bibnamefont{Sunyaev}},
  \bibinfo{journal}{Astron. Astrophys.} \textbf{\bibinfo{volume}{444}},
  \bibinfo{pages}{821} (\bibinfo{year}{2005}), \eprint{[astro-ph/0411550]}.

\bibitem[{\citenamefont{Corbet et~al.}(2007)}]{Corbet:2007vn}
\bibinfo{author}{\bibfnamefont{R.}~\bibnamefont{Corbet}} \bibnamefont{et~al.}
  (\bibinfo{collaboration}{Swift BAT}), \bibinfo{journal}{Prog. Theor. Phys.
  Suppl.} \textbf{\bibinfo{volume}{169}}, \bibinfo{pages}{200}
  (\bibinfo{year}{2007}), \eprint{[astro-ph/0703274]}.

\bibitem[{\citenamefont{Russell et~al.}(2013)}]{Russell:2013jva}
\bibinfo{author}{\bibfnamefont{D.~M.} \bibnamefont{Russell}}
  \bibnamefont{et~al.}, \bibinfo{journal}{Astrophys. J.}
  \textbf{\bibinfo{volume}{768}}, \bibinfo{pages}{L35} (\bibinfo{year}{2013}),
  \eprint{[arXiv:1304.3510]}.

  \bibitem[{\citenamefont{Corral-Santana
  et~al.}(2016)\citenamefont{Corral-Santana, Casares, Mu\~noz-Darias, Bauer,
  Martinez-Pais, and Russell}}]{Corral-Santana:2015fud}
\bibinfo{author}{\bibfnamefont{J.~M.} \bibnamefont{Corral-Santana}},
  \bibinfo{author}{\bibfnamefont{J.}~\bibnamefont{Casares}},
  \bibinfo{author}{\bibfnamefont{T.}~\bibnamefont{Mu\~noz-Darias}},
  \bibinfo{author}{\bibfnamefont{F.~E.} \bibnamefont{Bauer}},
  \bibinfo{author}{\bibfnamefont{I.~G.} \bibnamefont{Martinez-Pais}},
  \bibnamefont{and} \bibinfo{author}{\bibfnamefont{D.~M.}
  \bibnamefont{Russell}}, \bibinfo{journal}{Astron. Astrophys.}
  \textbf{\bibinfo{volume}{587}}, \bibinfo{pages}{A61} (\bibinfo{year}{2016}),
  \eprint{[arXiv:1510.08869]}.

\bibitem[{\citenamefont{Bogomazov}(2016)}]{Bogomazov:2016cei}
\bibinfo{author}{\bibfnamefont{A.~I.} \bibnamefont{Bogomazov}}
  (\bibinfo{year}{2016}), \eprint{[arXiv:1607.03358]}.

\bibitem[{\citenamefont{Abbott et~al.}(2016{\natexlab{b}})}]{Abbott:2016blz}
\bibinfo{author}{\bibfnamefont{B.~P.} \bibnamefont{Abbott}}
  \bibnamefont{et~al.} (\bibinfo{collaboration}{Virgo, LIGO Scientific}),
  \bibinfo{journal}{Phys. Rev. Lett.} \textbf{\bibinfo{volume}{116}},
  \bibinfo{pages}{061102} (\bibinfo{year}{2016}{\natexlab{b}}),
  \eprint{[arXiv:1602.03837]}.

\bibitem[{\citenamefont{Abbott et~al.}(2016{\natexlab{c}})}]{Abbott:2016nmj}
\bibinfo{author}{\bibfnamefont{B.~P.} \bibnamefont{Abbott}}
  \bibnamefont{et~al.} (\bibinfo{collaboration}{Virgo, LIGO Scientific}),
  \bibinfo{journal}{Phys. Rev. Lett.} \textbf{\bibinfo{volume}{116}},
  \bibinfo{pages}{241103} (\bibinfo{year}{2016}{\natexlab{c}}),
  \eprint{]arXiv:1606.04855]}.

\bibitem[{\citenamefont{Abbott
  et~al.}(2016{\natexlab{d}})}]{TheLIGOScientific:2016pea}
\bibinfo{author}{\bibfnamefont{B.~P.} \bibnamefont{Abbott}}
  \bibnamefont{et~al.} (\bibinfo{collaboration}{Virgo, LIGO Scientific})
  (\bibinfo{year}{2016}{\natexlab{d}}), \eprint{[arXiv:1606.04856]}.

\bibitem{Gerosa:2017kvu} 
  D.~Gerosa and E.~Berti,
  Phys.\ Rev.\ D {\bf 95}, no. 12, 124046 (2017)
  [arXiv:1703.06223 [gr-qc]].

\bibitem{Fishbach:2017dwv} 
  M.~Fishbach, D.~E.~Holz and B.~Farr,
  Astrophys.\ J.\  {\bf 840}, no. 2, L24 (2017)
  [arXiv:1703.06869 [astro-ph.HE]].
    
    \bibitem{Miller:2003sc} 
  M.~C.~Miller and E.~J.~M.~Colbert,
  Int.\ J.\ Mod.\ Phys.\ D {\bf 13}, 1 (2004)
  [astro-ph/0308402].
  
\bibitem{Berghea:2008rc} 
  C.~T.~Berghea, K.~A.~Weaver, E.~J.~M.~Colbert and T.~P.~Roberts,
  Astrophys.\ J.\  {\bf 687}, 471 (2008)
  [arXiv:0807.1547 [astro-ph]].
 

 \bibitem[Farrell et al.(2009)]{2009Natur.460...73F} S.~A.~Farrell, N.~A.~Webb, D.~Barret, O.~Godet and J.~M.~Rodrigues,  Nature {\bf 460}, 73 (2009) [arXiv:1001.0567].
 
 \bibitem[Davis et al.(2011)]{2011ApJ...734..111D} S.~W.~Davis, R.~Narayan, Y.~Zhu et al.\ Astrophys.\ J.\  {\bf 734}   111 (2011) [arXiv:1104.2614].
 
 \bibitem{Godet:2014bga} 
  O.~Godet, J.~C.~Lombardi, F.~Antonini, N.~A.~Webb, D.~Barret, J.~Vingless and M.~Thomas,
  Astrophys.\ J.\  {\bf 793}, no. 2, 105 (2014)
  [arXiv:1408.1819 [astro-ph.HE]].
 
   \bibitem{Pasham:2015tca} 
  D.~R.~Pasham, T.~E.~Strohmayer and R.~F.~Mushotzky,
  Nature {\bf 513}, 74 (2014)
  [arXiv:1501.03180 [astro-ph.HE]].
  
\bibitem{Mezcua} M.~Mezcua, Int.\ J.\ of Mod.\ Phy.\ D, {\bf 26}, 1730021 (2017) [arXiv:1705.09667].

\bibitem{Bachetti:2014qsa} 
  M.~Bachetti {\it et al.},
  Nature {\bf 514}, 202 (2014)
  [arXiv:1410.3590 [astro-ph.HE]].
  
  \bibitem{Bachetti:2015pwa} 
  M.~Bachetti,
  Astron.\ Nachr.\  {\bf 337}, no. 4/5, 349 (2017)
  [arXiv:1510.05565 [astro-ph.HE]].

\bibitem[Graham \& Scott(2013)]{2013ApJ...764..151G} A.~W.~Graham and N.~Scott, Astrophys.\ J.\  {\bf 764} , 151 (2013) [arXiv:1211.3199].

\bibitem{Kiziltan} B.~Kiziltan, H.~Baumgardt and A.~Loeb, Nature {\bf 542}, 203 (2017) [arXiv:1702.02149]. 

  
  \bibitem{Freire:2017mgu} 
  P.~C.~C.~Freire {\it et al.},
  Mon.\ Not.\ Roy.\ Astron.\ Soc.\  {\bf 471}, no. 1, 857 (2017)
  [arXiv:1706.04908 [astro-ph.HE]].
 
   \bibitem{Perera:2017jrk} 
  B.~B.~P.~Perera {\it et al.},
  Mon.\ Not.\ Roy.\ Astron.\ Soc.\  {\bf 468}, no. 2, 2114 (2017)
  [arXiv:1705.01612 [astro-ph.HE]].

 
\bibitem{Silk:2017yai} 
  J.~Silk,
  Astrophys.\ J.\  {\bf 839}, no. 1, L13 (2017)
  [arXiv:1703.08553 [astro-ph.GA]].

 \bibitem{Ebisuzaki:2001qm} 
  T.~Ebisuzaki {\it et al.},
  Astrophys.\ J.\  {\bf 562}, L19 (2001)
  [astro-ph/0106252].
 
 \bibitem{Brown:2006pj} 
  D.~A.~Brown, H.~Fang, J.~R.~Gair, C.~Li, G.~Lovelace, I.~Mandel and K.~S.~Thorne,
  Phys.\ Rev.\ Lett.\  {\bf 99}, 201102 (2007)
  [gr-qc/0612060].
  
\bibitem{Smith:2013mfa} 
  R.~J.~E.~Smith, I.~Mandel and A.~Vechhio,
  Phys.\ Rev.\ D {\bf 88}, no. 4, 044010 (2013)
  [arXiv:1302.6049 [astro-ph.HE]].

  \bibitem{Mandel:2008bc} 
  I.~Mandel and J.~R.~Gair,
  Class.\ Quant.\ Grav.\  {\bf 26}, 094036 (2009)
  [arXiv:0811.0138 [gr-qc]].
  
   \bibitem{AmaroSeoane:2007aw} 
  P.~Amaro-Seoane, J.~R.~Gair, M.~Freitag, M.~Coleman Miller, I.~Mandel, C.~J.~Cutler and S.~Babak,
  Class.\ Quant.\ Grav.\  {\bf 24}, R113 (2007)
  [astro-ph/0703495].
  
   \bibitem{Gair:2010yu} 
  J.~R.~Gair, C.~Tang and M.~Volonteri,
  Phys.\ Rev.\ D {\bf 81}, 104014 (2010)
  [arXiv:1004.1921 [astro-ph.GA]].
  
\bibitem{Flanagan:1997sx} 
  E.~E.~Flanagan and S.~A.~Hughes,
  Phys.\ Rev.\ D {\bf 57}, 4535 (1998)
  [gr-qc/9701039].


\bibitem{Berti:2005ys} 
  E.~Berti, V.~Cardoso and C.~M.~Will,
  Phys.\ Rev.\ D {\bf 73}, 064030 (2006)
  [gr-qc/0512160].

  \bibitem{Fregeau:2006yz} 
  J.~M.~Fregeau, S.~L.~Larson, M.~C.~Miller, R.~W.~O'Shaughnessy and F.~A.~Rasio,
  Astrophys.\ J.\  {\bf 646}, L135 (2006)
  [astro-ph/0605732].
  
  
\bibitem{Shinkai:2016xya} 
  H.~A.~Shinkai, N.~Kanda and T.~Ebisuzaki,
  Astrophys.\ J.\  {\bf 835}, no. 2, 276 (2017)
  [arXiv:1610.09505 [astro-ph.GA]].
  
\bibitem{Abbott:2017iws} 
  B.~P.~Abbott {\it et al.} [LIGO Scientific and Virgo Collaborations],
  Phys.\ Rev.\ D {\bf 96}, no. 2, 022001 (2017)
  [arXiv:1704.04628 [gr-qc]].
  
  \bibitem{Fishbach:2017zga} 
  M.~Fishbach and D.~E.~Holz,
  Astrophys.\ J.\  {\bf 851}, no. 2, L25 (2017)
  [arXiv:1709.08584 [astro-ph.HE]].
  
  \bibitem{CalderonBustillo:2017skv} 
  J.~Calder\'on Bustillo, F.~Salemi, T.~Dal Canton and K.~P.~Jani,
  Phys.\ Rev.\ D {\bf 97}, no. 2, 024016 (2018)
  [arXiv:1711.02009 [gr-qc]].
  
\bibitem{Gair:2010dx} 
  J.~R.~Gair, I.~Mandel, M.~C.~Miller and M.~Volonteri,
  Gen.\ Rel.\ Grav.\  {\bf 43}, 485 (2011)
  [arXiv:0907.5450].
  
  
  \bibitem{AtakanGurkan:2003hm} 
  M.~Atakan Gurkan, M.~Freitag and F.~A.~Rasio,
  Astrophys.\ J.\  {\bf 604}, 632 (2004)
  [astro-ph/0308449].

  \bibitem{PortegiesZwart:2004ggg} 
  S.~F.~Portegies Zwart, H.~Baumgardt, P.~Hut, J.~Makino and S.~L.~W.~McMillan,
  Nature {\bf 428}, 724 (2004)
  [astro-ph/0402622].
  
  \bibitem{Sakurai:2017opi} 
  Y.~Sakurai, N.~Yoshida, M.~S.~Fujii and S.~Hirano,
  Mon.\ Not.\ Roy.\ Astron.\ Soc.\  {\bf 472}, no. 2, 1677 (2017)
  [arXiv:1704.06130 [astro-ph.GA]].
  
  \bibitem{Miller:2001ez} 
  M.~C.~Miller and D.~P.~Hamilton,
  Mon.\ Not.\ Roy.\ Astron.\ Soc.\  {\bf 330}, 232 (2002)
  [astro-ph/0106188].
  
  \bibitem{PortegiesZwart:1999nm} 
  S.~F.~Portegies Zwart and S.~McMillan,
  Astrophys.\ J.\  {\bf 528}, L17 (2000)
  [astro-ph/9910061].
  
\bibitem{Rodriguez:2015oxa} 
  C.~L.~Rodriguez, M.~Morscher, B.~Pattabiraman, S.~Chatterjee, C.~J.~Haster and F.~A.~Rasio,
  Phys.\ Rev.\ Lett.\  {\bf 115}, no. 5, 051101 (2015)
  [arXiv:1505.00792].

\bibitem{Samsing:2017xmd} 
  J.~Samsing,
  arXiv:1711.07452 [astro-ph.HE].
  
\bibitem{Gondan:2017wzd} 
  L.~Gond\'an, B.~Kocsis, P.~Raffai and Z.~Frei,
  arXiv:1711.09989 [astro-ph.HE].

\bibitem{Kovetz:2016kpi} 
  E.~D.~Kovetz, I.~Cholis, P.~C.~Breysse and M.~Kamionkowski,
  Phys.\ Rev.\ D {\bf 95}, no. 10, 103010 (2017)
  [arXiv:1611.01157 [astro-ph.CO]].

\bibitem{LALwebsite} 
  {\tt   https://wiki.ligo.org/DASWG/LALSuite}.
  
  \bibitem{Matsubayashi} 
  T.~Matsubayashi, H.~A.~Shinkai and T.~Ebisuzaki,
  Astrophys.\ J.\  {\bf 614}, no. 2, 864 (2004)
  

  \bibitem{Fragione:2017blf} 
  G.~Fragione, I.~Ginsburg and B.~Kocsis,
  arXiv:1711.00483 [astro-ph.GA].
  
\bibitem{NASAwebsite} 
  {\tt https://heasarc.gsfc.nasa.gov/W3Browse 
  /star-catalog/globclust.html}.

\bibitem{King1962} 
  I.~King 
  \aj \, {\bf 67}, 471K, 1962.

\bibitem{Baumgardt:2004tp} 
  H.~Baumgardt, J.~Makino and P.~Hut,
  Astrophys.\ J.\  {\bf 620}, 238 (2005)
  [astro-ph/0410597].

\bibitem{Kroupa:2000iv} 
  P.~Kroupa,
  Mon.\ Not.\ Roy.\ Astron.\ Soc.\  {\bf 322}, 231 (2001) [astro-ph/0009005].

\bibitem{QuinlanShapiro1989}
 G.~D.~Quinlan and S.~L.~Shapiro,  Astrophys. J. {\bf 343}, 725 (1989).

\bibitem{MouriTaniguchi2002}
H.~Mouri and Y.~Taniguchi, \; Astrophys. J. 566, {\bf L17}  (2002).

\bibitem{PetersMathews1963}
P.~C.~Peters and J.~Mathews, Phys. Rev. {\bf 131}, 435  (1963).

\bibitem{Peters1964}
P.~C.~Peters, Phys. Rev. {\bf 136},  B1224 (1964).

\bibitem{Cholis:2016kqi} 
  I.~Cholis, E.~D.~Kovetz, Y.~Ali-Ha\"imoud, S.~Bird, M.~Kamionkowski, J.~B.~Mu\~noz and A.~Raccanelli,
  Phys.\ Rev.\ D {\bf 94}, no. 8, 084013 (2016)
  [arXiv:1606.07437].
  
\bibitem{Morscher:2014doa} 
  M.~Morscher, B.~Pattabiraman, C.~Rodriguez, F.~A.~Rasio and S.~Umbreit,
  Astrophys.\ J.\  {\bf 800}, no. 1, 9 (2015)
 [arXiv:1409.0866 [astro-ph.GA]].

\bibitem{2016MNRAS.463.2109R} 
Rodriguez, C.~L., Morscher, M., Wang, L., et al.
2016, \mnras, 463, 2109 [arXiv:1601.04227].

\bibitem{mandel08}
I.~Mandel, D.~A.~Brown, J.~R.~Gair, M.~C.~Miller \; Astrophys. J. {\bf 681}, 1431 (2008) [arXiv:0705.0285].

  \bibitem{Haster:2016ewz} 
  C.~J.~Haster, F.~Antonini, V.~Kalogera and I.~Mandel,
  Astrophys.\ J.\  {\bf 832}, no. 2, 192 (2016)
  [arXiv:1606.07097].

  \bibitem{Antonini:2013tea} 
  F.~Antonini, N.~Murray and S.~Mikkola,
  Astrophys.\ J.\  {\bf 781}, 45 (2014)
  [arXiv:1308.3674 [astro-ph.HE]].
  
  \bibitem{Samsing:2013kua} 
  J.~Samsing, M.~MacLeod and E.~Ramirez-Ruiz,
  Astrophys.\ J.\  {\bf 784}, 71 (2014)
  [arXiv:1308.2964 [astro-ph.HE]].

\bibitem{Kozai:1962zz} 
  Y.~Kozai,
  Astron.\ J.\  {\bf 67}, 591 (1962).
  
  \bibitem{Miller:2002pg} 
  M.~C.~Miller and D.~P.~Hamilton,
  Astrophys.\ J.\  {\bf 576}, 894 (2002)
 [astro-ph/0202298].
  
  \bibitem{OLeary:2005vqo} 
  R.~M.~O'Leary, F.~A.~Rasio, J.~M.~Fregeau, N.~Ivanova and R.~W.~O'Shaughnessy,
  Astrophys.\ J.\  {\bf 637}, 937 (2006)
 [astro-ph/0508224].

\bibitem{Hopman:2005hm} 
  C.~Hopman and S.~F.~Portegies Zwart,
  Mon.\ Not.\ Roy.\ Astron.\ Soc.\  {\bf 363}, L56 (2005) 
  [astro-ph/0506181].

 \bibitem{MacLeod:2015bpa} 
  M.~MacLeod, M.~Trenti and E.~Ramirez-Ruiz,
  Astrophys.\ J.\  {\bf 819}, no. 1, 70 (2016)
  [arXiv:1508.07000 [astro-ph.HE]].
  
    \bibitem{Miller:2002vg} 
  M.~C.~Miller,
  Astrophys.\ J.\  {\bf 581}, 438 (2002)
  [astro-ph/0206404].

\bibitem{HolleyBockelmann:2007eh} 
  K.~Holley-Bockelmann, K.~Gultekin, D.~Shoemaker and N.~Yunes,
  Astrophys.\ J.\  {\bf 686}, 829 (2008)
  [arXiv:0707.1334 [astro-ph]].

  \bibitem{Kovetz:2017rvv} 
  E.~D.~Kovetz,
  Phys.\ Rev.\ Lett.\  {\bf 119}, no. 13, 131301 (2017)
  [arXiv:1705.09182 [astro-ph.CO]].

 \bibitem{Abbott:2017vtc} 
  B.~P.~Abbott {\it et al.} [LIGO Scientific and VIRGO Collaborations],
  Phys.\ Rev.\ Lett.\  {\bf 118}, no. 22, 221101 (2017)
  [arXiv:1706.01812 [gr-qc]].
 
\bibitem{TheLIGOScientific:2016wyq} 
  B.~P.~Abbott {\it et al.} [LIGO Scientific and Virgo Collaborations],
  Phys.\ Rev.\ Lett.\  {\bf 116}, no. 13, 131102 (2016)
  [arXiv:1602.03847 [gr-qc]].
  
\bibitem{Cholis:2016xvo} 
  I.~Cholis,
  JCAP {\bf 1706}, no. 06, 037 (2017)
  [arXiv:1609.03565 [astro-ph.HE]].
  
  \bibitem{Abbott:2017xzg} 
  B.~P.~Abbott {\it et al.} [LIGO Scientific and Virgo Collaborations],
  arXiv:1710.05837 [gr-qc].
  
  
  \bibitem{Madau:2014bja} 
  P.~Madau and M.~Dickinson,
  Ann.\ Rev.\ Astron.\ Astrophys.\  {\bf 52}, 415 (2014)
  [arXiv:1403.0007 [astro-ph.CO]].
  
  \bibitem{Spera:2016slz} 
  M.~Spera, N.~Giacobbo and M.~Mapelli,
  Mem.\ Soc.\ Ast.\ It.\  {\bf 87}, 575 (2016)
  [arXiv:1606.03349 [astro-ph.SR]].
  
  \bibitem{Spera:2017fyx} 
  M.~Spera and M.~Mapelli,
  Mon.\ Not.\ Roy.\ Astron.\ Soc.\  {\bf 470}, no. 4, 4739 (2017)
  [arXiv:1706.06109 [astro-ph.SR]].
        

  \bibitem{Chen:2017wpg} 
  H.~Y.~Chen, D.~E.~Holz, J.~Miller, M.~Evans, S.~Vitale and J.~Creighton,
  arXiv:1709.08079 [astro-ph.CO].
  
  
\bibitem{Veitch:2014wba} 
  J.~Veitch {\it et al.},
  Phys.\ Rev.\ D {\bf 91}, no. 4, 042003 (2015)
  [arXiv:1409.7215 [gr-qc]].

\bibitem{Smith:2016qas} 
  R.~Smith, S.~E.~Field, K.~Blackburn, C.~J.~Haster, M.~P\"urrer, V.~Raymond and P.~Schmidt,
  Phys.\ Rev.\ D {\bf 94}, no. 4, 044031 (2016)
  [arXiv:1604.08253 [gr-qc]].
  
    \bibitem{Aasi:2013wya} 
  B.~P.~Abbott {\it et al.} [LIGO Scientific and VIRGO Collaborations],
  Living Rev.\ Rel.\  {\bf 19}, 1 (2016)
  [arXiv:1304.0670 [gr-qc]].

    \bibitem{GWDistCalcwebsite} 
  {\tt http://gwc.rcc.uchicago.edu/}

  \bibitem{Haster:2015cnn} 
  C.~J.~Haster, Z.~Wang, C.~P.~L.~Berry, S.~Stevenson, J.~Veitch and I.~Mandel,
  Mon.\ Not.\ Roy.\ Astron.\ Soc.\  {\bf 457}, no. 4, 4499 (2016)
  [arXiv:1511.01431 [astro-ph.HE]].

\bibitem{Park:2017zgj} 
  D.~Park, C.~Kim, H.~M.~Lee, Y.~B.~Bae and K.~Belczynski,
  Mon.\ Not.\ Roy.\ Astron.\ Soc.\  {\bf 469}, no. 4, 4665 (2017)
  [arXiv:1703.01568 [astro-ph.HE]].


  \bibitem{ET}
  Einstein Telescope, design at: http://www.et-gw.eu/
  
\bibitem[{\citenamefont{Sathyaprakash et~al.}(2012)}]{Sathyaprakash:2012jk}
\bibinfo{author}{\bibfnamefont{B.}~\bibnamefont{Sathyaprakash}}
  \bibnamefont{et~al.}, \bibinfo{journal}{Class. Quant. Grav.}
  \textbf{\bibinfo{volume}{29}}, \bibinfo{pages}{124013}
  (\bibinfo{year}{2012}), \eprint{[arXiv:1206.0331]}.
  
  \bibitem{Evans:2016mbw} 
  B.~P.~Abbott {\it et al.} [LIGO Scientific Collaboration],
  Class.\ Quant.\ Grav.\  {\bf 34}, 044001 (2017)
  [arXiv:1607.08697].

  
  \bibitem{Seto:2001qf} 
  N.~Seto, S.~Kawamura and T.~Nakamura,
  Phys.\ Rev.\ Lett.\  {\bf 87}, 221103 (2001)
  [astro-ph/0108011].

  \bibitem[{\citenamefont{Amaro-Seoane et~al.}(2012)}]{AmaroSeoane:2012je}
\bibinfo{author}{\bibfnamefont{P.}~\bibnamefont{Amaro-Seoane}}
  \bibnamefont{et~al.}, \bibinfo{journal}{Class. Quant. Grav.}
  \textbf{\bibinfo{volume}{29}}, \bibinfo{pages}{124016}
  (\bibinfo{year}{2012}), \eprint{[arXiv:1202.0839]}.


 \bibitem[Amaro-Seoane et al.(2017)]{2017arXiv170200786A} P.~Amaro-Seoane, H.~Audley, S.~Babak et al., arXiv:1702.00786 
  
  
\end{thebibliography}
\end{document}